\documentclass[epj]{svjour}
\usepackage{latexsym}
\RequirePackage{graphicx}
\RequirePackage{mathptmx}      
\usepackage{amsmath}
\usepackage[bottom]{footmisc}
\usepackage{float}
\begin{document}
\title{{De-channeling in terms of instantaneous transition rates}
\subtitle{Computer simulations for 855 MeV electrons at (110) planes of diamond
}
}
\author{H. Backe
}                     

\institute{Institute for Nuclear Physics of Johannes Gutenberg-University, Johann-Joachim-Becher-Weg 45, 55128 Mainz, Germany}
%
%
\abstract{
Monte-Carlo simulation calculation have been performed for 855 MeV electrons channeling in (110) planes of a diamond single crystal. The continuum potential picture has been utilized. Both, the transverse potential and the angular distributions of the scattered electrons at screened atoms are based on the Doyle-Turner scattering factors which were extrapolated with the functional dependence of the Moli\`{e}re representation to large momentum transfers. Scattering cross-sections at bound electrons have been derived for energies less than 30 keV from the double differential cross-section as function of both, energy and momentum transfer, taking into account also longitudinal and transverse excitations. For energies above 30 keV the M{\o}ller cross-section is used. The dynamics of the particle in the continuum transverse potential has been described classically. Results of the channeling process are presented in terms of instantaneous transition rates as function of the penetration depth, indicating that channeling can be described by a single exponential function only after the equilibration phase has been reached after about 15 $\mu$m. As a byproduct, improved drift and diffusion coefficients entering the Fokker-Planck equation have been derived with which its predictive power can be improved.
} 
\maketitle
\section{Introduction}\label{introduction}
There exists since long considerable interest in the channeling process of ultra-relativistic electrons and positrons at planes of a single crystal. Of particular interest is the emission of undulator-like radiation in periodically bent crystals aiming in the construction of compact radiation sources in the MeV range and beyond, see e.g. Korol et al. \cite{KorS14}. However, collisions of the leptons with atoms and their electrons comprising the crystallographic planes lead to a de-channeling process being in conflict with the envisaged application. It is, therefore, of utmost importance to understand the channeling as well as the de- and also re-channeling processes in detail, in particular also for electrons.

While "de-channeling" is a well defined technical term, the de-channeling length is not. Indeed, a particle entering a straight channel needs some time, or a certain length interval, to equilibrate. Therefore, a constant de-channeling rate, leading to an exponential decrease of the channel occupation does generally not exist. Reliable experimental information on the rate distribution as function of the penetration depth of the particle in the crystal is rare, or does not exist at all. However, such information is required to determine the channeling efficiency, i.e., that fraction of particles which follow the exponential decay law. This paper intends to contribute to this task via Monte Carlo simulations. This way, also the re-channeling process has been studied, including the time when re-channeling happens, the number of planes the electron passes until it finally re-channels, the probability distribution, and the efficiency of the re-channeling process.

For many years channeling of ultra-relativistic particles in single crystals was described in the continuum potential picture introduced by Lindhard \cite{Lin65}. In particular, the Fokker-Planck equation with which the de-channeling process can be calculated is based on it, see e.g. the papers of Backe et al. \cite{BacL08,Bac18} and references cited therein. Objections against the reliability of the solutions of the Fokker-Planck equation have been formulated by Tikhomirov \cite[ch. 3.4]{Tik17} but it has not been investigated whether its predictive power can be improved, or not. This paper will contribute also to this subject, in particular it focuses on a better approximation of the Kitagawa-Ohtsuki drift and diffusion coefficients \cite{KitO73} still in use.

If experimental results do not exist, insight into the addressed issues can be obtained by Monte Carlo simulation calculations. There exist detailed considerations with sophisticated codes, see e.g. Pavlov et al. \cite{PavK20} and references cited therein, and the continuum potential picture seems to be outdated. However, a detailed comparison between the former and the latter with a discussion of advantages and disadvantages does not exist. This issue will be addressed in passing.

The continuum potential picture has the advantage that channeling can be classically understood for ultra-relativistic particles in a rather intuitive manner. In particular, for ultra-relativistic leptons the dynamics can be treated classically. Deficiencies which were already pointed out by Lindhard, namely that the potential has fluctuations in the Angstrom scale should have only little impact. A lepton moving in forward direction at angles in the order sub milli-radiants feels during its passage the potential of thousands of single atoms. It is therefore expected that the potential can be described rather precisely in such a collective approach. Moreover, it can be considered as decoupled from hard scattering events at atoms and bound electrons since those interactions are rare.

Channeling in the continuum potential picture was treated by a number of authors, for an overview see, e.g., Korol et al. \cite[Chapter 2]{KorS21}. In particular, the DYNECHARM++ code of Bagli and Guidi \cite{BagG13}, and CRYSTALRAD of Sytov et al. \cite{SytT19} should be mentioned. In this work yet another code was developed which is restricted to the above mentioned problems, however, with the possibility to investigate the sensitivity on special interactions like low energy plasmon excitations.

The paper is organized as follows. Sections 2-4 summarize the theoretical background. In section \ref{particle dynamics} basic definitions and the particle dynamics are described. In section \ref{interaction with atoms} the Doyle-Turner representation of the electron scattering factors at screened atoms is introduced with a modification into the Moli\`{e}re approach. The latter has the advantage that it avoids a cutoff at large momentum transfers. On the basis of this data set the electric potential, the electron density, and collision cross-sections are calculated. The latter finally leads to the scattering distribution needed for the Monte Carlo simulations. In section \ref{interaction with electrons} and in the appendix \ref{appendix A} the electron-electron interaction is treated. At energy losses larger than about 30 keV the M{\o}ller cross section has been applied. For less than 30 keV models of Ashley \cite{Ash91} and Fernández-Varea \cite{FerS05} were utilized leading to the double differential cross-sections as function of both, the energy and momentum transfer to bound electrons from which the electron-electron scattering distribution was derived. In section \ref{calculations} preparatory calculations are described which include the test of the model using the example of an amorphous carbon foil, and the calculation of the initial probability distribution. In section \ref{results} results of the Monte Carlo simulations for channeling and re-channeling of 855 MeV electrons in straight and bent (110) planes of diamond are presented, and for the latter compared with published results obtained with the MBN explorer package. In section \ref{sectionFokkerPlanck} drift and diffusion coefficients, which enter the Fokker-Planck equation, were calculated for both, scattering at atoms and separately at electrons, and the impact on its predicting power is discussed. The paper closes with a conclusion in section \ref{conclusions}.

\section{Particle dynamics}\label{particle dynamics}
For a plane crystal a laboratory coordinate system $(x, y, z)$ has been chosen with the $z$ axis located in the crystallographic plane, the horizontal $x$ axis perpendicular to it, and the vertical $y$ axis such that a right handed coordinate system results. The planes of a bent crystal are assumed to be circularly shaped with a constant bending radius $R$, and a concave shape with respect to the $z$ axis. The entrance angle into the crystal $\psi_0$ is the projected angle onto the (x,z) plane measured with respect to the $z$ axis. The rotation of the crystal around the $y$ axis has a negative sign for clockwise, and positive one for counter clockwise rotation. Analytically the following expressions were used:
\begin{equation}\label{bending1}
x = R \cdot \Big(\sqrt{1-(\sin\psi_0-z/R)^2}-\cos\psi_0~\Big)
\end{equation}

\begin{equation}\label{bending2}
\hspace{-1,3cm}
\frac{dx}{dz} = \vartheta_x = \frac{\sin\psi_0-z/R}{\sqrt{(1-(\sin\psi_0-z/R))^2}}
\end{equation}

\begin{equation}\label{bending3}
\frac{d^2x}{dz^2} = \frac{d \vartheta_x}{dz} \cong  -\frac{1}{R}.
\end{equation}

In the simulations it has been assumed that the particle moves freely a certain distance which will be randomly interrupted by hard collisions with lattice atoms and/or electrons. In the free movement the total energy is conserved while in a collision a sudden transverse energy transfer happens without a change of the $x$ coordinate. The calculations were performed in the $(x, E_\bot)$ phase space in a grid with step sizes $\Delta x \simeq$ 0.01 $\AA$ and exact transverse energy $E_\bot$ in the potential well. For the movement of a particle at position $x_k$ with transverse energy $E_{\bot,k}$ to $x_{k+1}~=~x_k + \Delta x_k$ a time $\Delta t_k$ is required. In the following the time is always multiplied by the particles velocity $v\simeq c$, and one obtains for the step size in $z$ direction
\begin{equation}
\Delta z_k=\Delta t_k~v=\int\limits_{x_k}^{x_k+\Delta x_k}\frac{\sqrt{p v/2}} {\sqrt{E_{\bot,k}-U(x)}}dx.
\label{deltaTstep}
\end{equation}
Here $p v = \beta\gamma m_e c^2$ is the momentum of the particle, with $\gamma=1/\sqrt{1-\beta^2}$ the Lorentz factor, $\beta=v/c$ and $m_e$ the electron rest mass. Despite the rather small integration interval $\Delta x_k$, the integral representation must be chosen to obtain reliable $\Delta z_k$, in particular at the turning points in the potential wall $U(x)$ since the denominator tends to zero there. In that case $\Delta x_k$ reduces to $\Delta x_k = \mid U^{-1}(E_{\bot,k})-x_k\mid$ and $\Delta z_k$ doubles. Eq. (\ref{deltaTstep}) is written for a bent crystal for which is
\begin{equation}
U(x)=u(x)-(pv/R)x,
\label{totalPotential}
\end{equation}
with $u(x)$ the potential of the plane crystal. However, this approach only holds for a constant bending radius $R$, i.e., not if the bending radius is a function of $z$.

At the distance $\Delta z_k$ a mean number of collisions happen, given by the total cross-section, from which according to the Poisson-statistics integer numbers are randomly generated. Thereafter, the individual scattering angle is randomly simulated for each scattering event with a procedure described below from which the transverse energy change is calculated.

\section{Interactions with atoms}\label{interaction with atoms}
\begin{figure}[tbh]
\centering
    \includegraphics[angle=0,scale=0.45,clip]{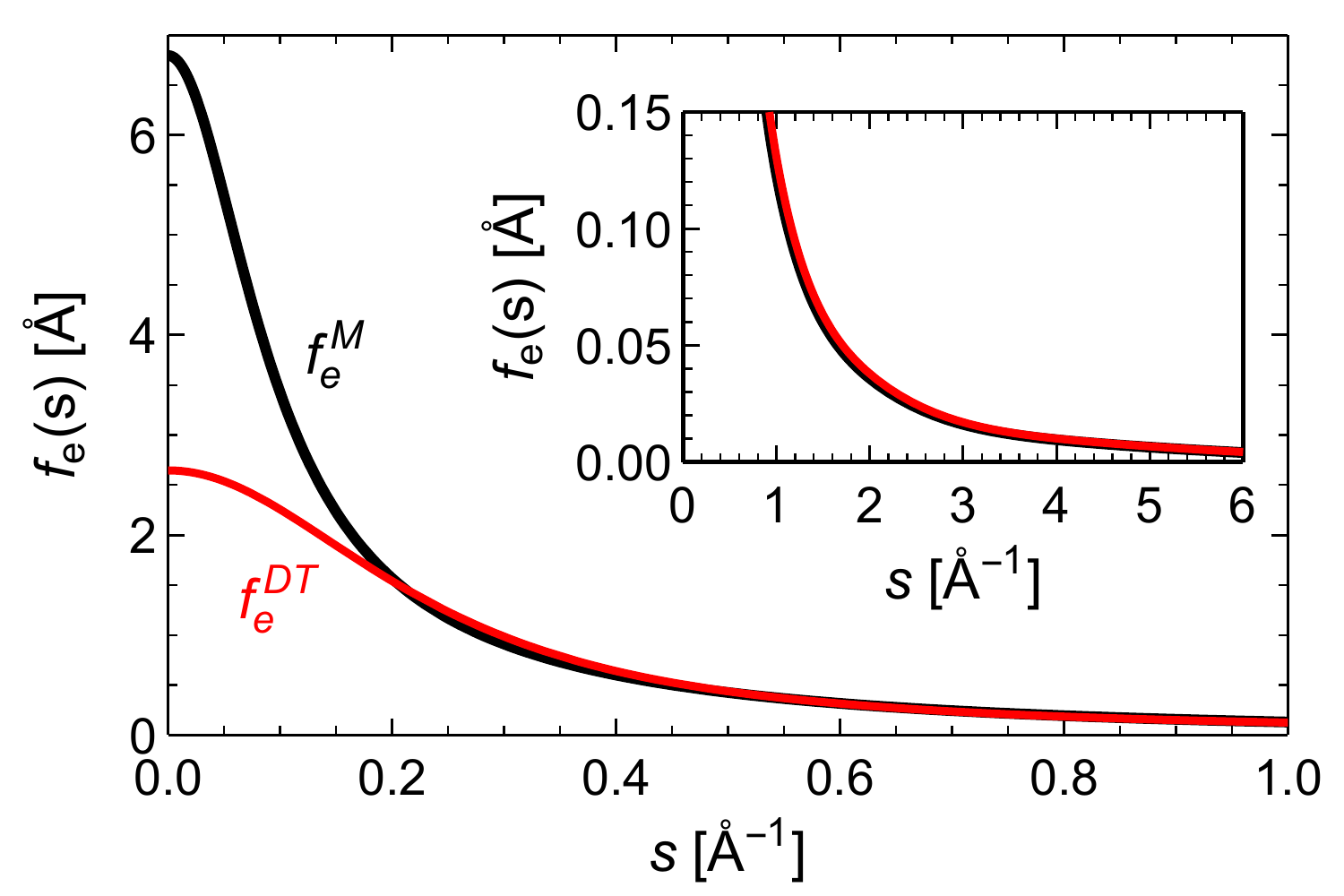}
\caption[]{Electron scattering factors for atomic carbon. The black full curve $f_e^{M}$ represents the Moli\`{e}re approximation \cite{Mol47} according to  Eq. (\ref{ScatterFactorM}), the red one $f_e^{DT}$ that one of Doyle and Turner \cite{DoyT67} in the 6-parameter representation of Eq. (\ref{ScatterFactorDT}) with values from Ref. \cite{ChoU99}.} \label{ScatteringFactors}
\end{figure}

For the calculation of the channeling potential, as well as for the also needed angular distribution of the scattered electrons, the electron scattering factors at screened atoms are required. There exist essentially two approaches to access them, that of Moli\`{e}re \cite{Mol47} and that of Doyle and Turner \cite{DoyT67}. Using the parameters
\begin{eqnarray}\label{MoliereParam}
\alpha^M &=& \{0.1,~~0.55,~~0.35\}\\
\beta^M &=& \{6.0,~~~1.2,~~~~~ 0.3\}/a_{TF}
\end{eqnarray}
with which Moli\`{e}re adapted the Thomas-Fermi function, a scattering factor
\begin{eqnarray}\label{ScatterFactorM}
f_e^{M} (s)=\frac{4\pi Z \alpha \hbar c}{a^3} \sum_{i=1}^{3}\frac{\alpha_i^M}{(4\pi s)^2+(\beta_i^M)^2}
\end{eqnarray}
results. It is compared  in Fig. \ref{ScatteringFactors} with the Doyle-Turner representation
\begin{eqnarray}\label{ScatterFactorDT}
f_e^{DT} (s)=\frac{2\pi a_0 \alpha \hbar c}{a^3} \sum_{i=1}^{6}a_i \exp(-b_i s^2)
\end{eqnarray}
using six pairs of parameters ($a_i,b_i$) which were quoted by Chouffani and \"{U}berall \cite{ChoU99}.
Here are $a_{TF}=0.8853 ~a_0 ~Z^{-1/3}$ the Thomas-Fermi screening factor, $a_0$ the Bohr radius, $Z$=6 for carbon, and $a$ = 3.5668 ${\AA}$ the lattice constant of diamond \cite{website:ioffe}. The quantity $s$ is related to the momentum transfer by $q=2p v/(\hbar c) \sin(\vartheta/2) = 4\pi s$, with $\vartheta$ the scattering angle.
Significant deviations up to a factor of 2.6 can be observed for $s<0.2~\AA^{-1}$ while for larger values they are less than 8\%. Nevertheless, all calculations in this paper have been performed in the Moli\`{e}re approximation, however, with the modified parameter set
\begin{eqnarray}\label{MoliereParamMod}
\alpha &=& \{0.09567,~2.9992\cdot10^{-5},~0.90430\}\nonumber \\
\beta &=& \{9.61747,~~~~~~~1.2,~~~~~~~~~~~0.73393\}/a_{TF}
\end{eqnarray}
which approximates the Doyle-Turner representation to better than 8\% in the full range of $0 \leq s/{\AA} \leq 6$. The reason is that $f_e^M(s)$ asymptotes to a Lorentzian rather than to a Gaussian, avoiding the unrealistic cut-off for large momentum transfers.
\subsection{Calculation of the continuum potential}\label{continuum potential}
\begin{figure}[tbh]
\centering
    \includegraphics[angle=0,scale=0.5,clip]{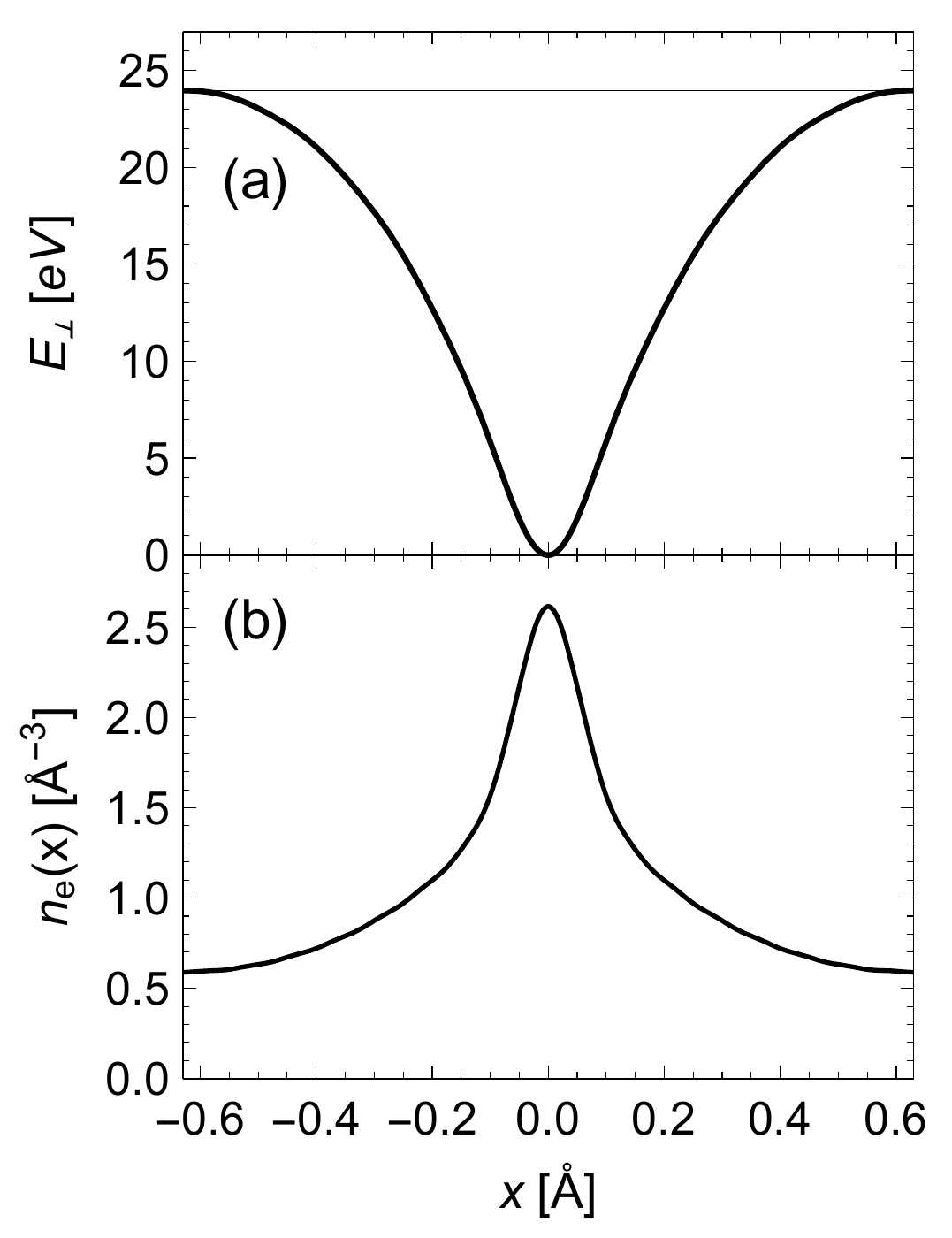}
\caption[]{(a) Potential of the (110) plane for a plane diamond crystal. The potential minimum has been normalized to zero, i.e. the continuum border is located at $E_{\bot,0}$ = $u_0$ = 23.957 eV. (b) Electron density $n_e(x)$ as function of the inter-planar distance coordinate $x$. The interplanar distance is $d_p~=~1.261~$ {\AA}, and the mean electron density for diamond $(1/d_p)\int_{-d_p/2}^{d_p/2} n_e(x) dx = Z\cdot (8/a^3)$ = 1.0578/${\AA}^3$.} \label{potential}
\end{figure}

The potential $u(x)$ has been calculated according to chapter 9.1 of the textbook of Baier et al. \cite{BaiK98} with the Fourier-expansion coefficients
\begin{eqnarray}\label{ScatterFactorMol}
G(q)=\frac{4\pi Z \alpha \hbar c}{a^3} \exp\Big(-\frac{u_1^2 q^2}{2}\Big) S(q) \sum_{i=1}^{3}\frac{\alpha_i}{q^2+\beta_i^2}
\label{FormFactor}
\end{eqnarray}
and $S(q)$ the structure factor. The one dimensional thermal vibration amplitude is $u_1$ = 0.04226 \AA ~\cite{SeaS91}. The potential $u(x)$ is shown in Fig.\ref{potential} (a), and in panel (b) the electron density $n_{el}(x)$ across the channel. The latter has been derived from the potential $u(x)$ with the aid of the one dimensional Poisson equation.

\subsection{Scattering distribution for atoms}\label{scattering with atoms}
Collision with atoms will be described in the screened potential with the modified Moli\`{e}re parameters of Eq. (\ref{MoliereParamMod}). The differential cross-section reads
\begin{equation}
\frac{d\sigma^{(at)}}{d\Omega}(\vartheta)=4(Z\alpha)^2 \big(\frac{\hbar c}{pc}\big)^2 \bigg(\sum_{i=1}^{3}\frac{\alpha_i}{4 \sin^2(\vartheta/2)+(\beta_i\cdot\hbar c/pc)^2}\bigg)^2.
\label{differentialCrossSection}
\end{equation}
Eq. (\ref{differentialCrossSection}) can be derived from the Mott cross-section by replacing $1/q^2$ by
\begin{equation}
\frac{1-F(q)}{q^2}=\sum_{i=1}^{3}\frac{\alpha_i}{q^2+\beta_i^2},
\label{FormFactor}
\end{equation}
with $F(q)$ the atomic form-factor. At planar channeling we are interested in the differential scattering cross-section with respect to the $x$ coordinate. It is obtained from Eq. (\ref{differentialCrossSection}) for $\vartheta\ll 1$, and $\vartheta^2=\vartheta_x^2+\vartheta_y^2$ by integration over the $\vartheta_y$ coordinate and reads approximately for $\vartheta_x\ll 1$
\begin{eqnarray}\label{differentialCrossSectionX}
\frac{d\sigma^{(at)}}{d\vartheta_x}(\vartheta_x)\cong4(Z\alpha)^2 \big(\frac{\hbar c}{pc}\big)^2\times
\nonumber\\
& &\hspace{-3,0 cm}
\int_{-\infty}^\infty\bigg(\sum_{i=1}^{3}\frac{\alpha_i}{\vartheta_x^2+{\vartheta_y'}^2+ (\beta_i\cdot\hbar c/pc)^2}\bigg)^2 d{\vartheta_y}'.
\end{eqnarray}
From the total cross-section
\begin{equation}
\sigma_{tot}^{(at)}=\int_{0}^\pi \frac{d\sigma^{(at)}}{d\Omega}(\vartheta)2\pi\sin\vartheta d \vartheta\cong\int_{-\infty}^\infty\frac{d\sigma^{(at)}}{d\vartheta_x}(\vartheta_x) d\vartheta_x
\label{totalCrossSection}
\end{equation}
one obtains the mean number of collisions in an interval $\Delta z_k=\Delta t_k~v$ as
\begin{equation}
\Delta m_k=\frac{8}{a^3}\Delta z_k~\cdot\sigma_{tot}^{(at)} \frac{d_p}{\sqrt{2\pi}u_1} \exp(-x_k^2/2u_1^2).
\label{meanCollisions}
\end{equation}
Here $a$ = 3.5668 $\AA$ is the lattice constant of diamond \cite{website:ioffe}, $8/a^3$ the atomic number density, and $d_p=a/2\sqrt{2}$ = $1.261~{\AA}$ the inter-planar distance. Due to the distribution of the atoms comprising the planes by thermal vibrations, the mean number of collisions is also a function of the inter-planar coordinate $x_k$. The probability distribution for $m$ collisions at a mean $\Delta m_k$ is given by the Poisson-distribution function
\begin{equation}
P_m(k)=\frac{(\Delta m_k)^m}{m!}\exp(-\Delta m_k).
\label{PoissonDistributionAt}
\end{equation}
The underlying assumption for applying this Eq. (\ref{PoissonDistributionAt}) is statistical independence of the collisions which should be fulfilled if the particle direction does not coincide with a crystal axis. In the simulation $M_k = \mbox{Random}_{m}[P_{m}(k)]$ is randomly generated. For $M_k$ = 0 the change of the scattering angle is $\Delta\vartheta_{k}^{(at)}$ = 0, otherwise
\begin{equation}
\Delta\vartheta_{k}^{(at)} = \sum_{i=1}^{M_k} \mbox{Random}_{i,\theta_{x}}[P^{(at)}(\theta_{x})],
\label{changeScatteringAngle}
\end{equation}
is obtained by a repeated generation of random numbers from the normalized scattering distribution function $P^{(at)}(\theta_{x})$. The latter has been derived from the angular part of Eq. (\ref{differentialCrossSectionX}) after normalization with the total cross-section Eq. (\ref{totalCrossSection}). It is a rather involved function on the parameters $\alpha_i$ and $\beta_i$. Numerically one obtains
\begin{eqnarray} \label{normalScattProb}
\hspace{-0.5 cm}
P^{(at)}(\theta_x)  = \Big(\frac{1.68969 \cdot 10^{-8}+\theta_x^2}{(4.3114\cdot 10^{-11} + \theta_x^2)^{3/2}}
\nonumber\\
& &\hspace{-3.8 cm}
-\frac{3.56831 \cdot 10^{-12} +0.03096\cdot \theta_x^2}{(1.1526\cdot 10^{-10} + \theta_x^2)^{3/2}}
\nonumber\\
& &\hspace{-3.8 cm}
-\frac{6.98552 \cdot 10^{-9} + 0.96904\cdot \theta_x^2}{(7.4034\cdot 10^{-9} + \theta_x^2)^{3/2}}\Big)\Big/786.89.
\end{eqnarray}
\begin{figure}[tbh]
\vspace*{-0.4cm}
\centering
    \includegraphics[angle=0,scale=0.45,clip]{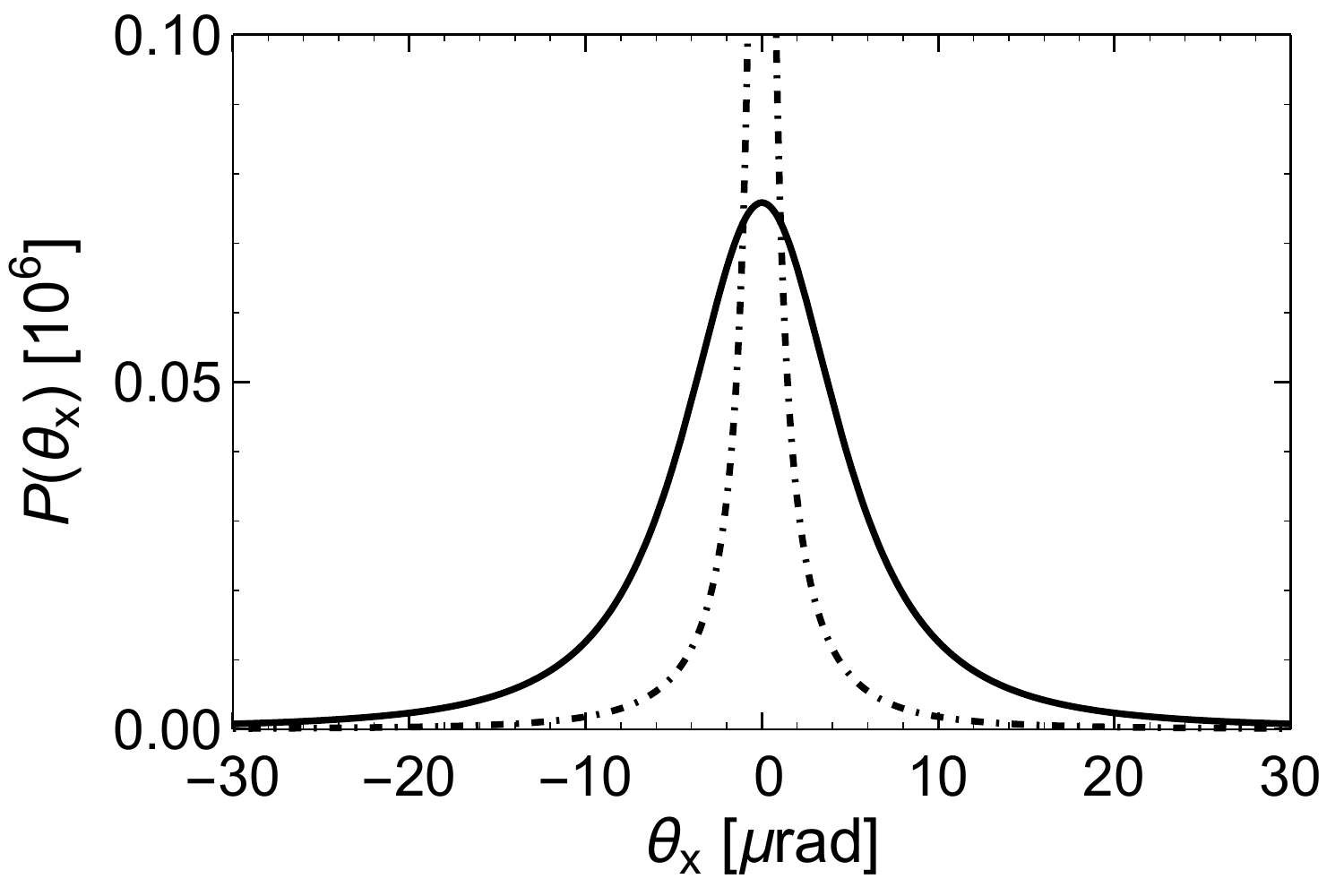}
\caption[]{Normalized atomic (full) and electronic (dot-dashed) scattering distributions $P^{(at)}(\theta_x)$ and $P^{(el)}(\theta_x)$, respectively, as function of the scattering angle $\theta_x$ for 855 MeV electrons on carbon atoms. The electronic distribution is very narrow extending vertically up to 3.73$\cdot 10^6$. For a logarithmic representation of the electronic distribution see Fig. \ref{ScatteringProbabilitiesEl}. The FWHM amount to 10.08 $\mu$rad (atomic), and 0.0752 $\mu$rad (electronic). Both distributions have long tails taken into account in the numerical simulation up to $\pm$ 0.02 rad (atomic) and $\pm$ 0.0345 rad (electronic). The root mean squared scattering angles are $<\theta_x^2>^{1/2}$ = 19.6 $\mu$rad (atomic), and 7.33 $\mu$rad (electronic).} \label{ProbabilityWnorm}
\end{figure}
The distribution function is shown in Fig. \ref{ProbabilityWnorm}, full curve. The total scattering cross-section according to Eq. (\ref{totalCrossSection}) is $\sigma_{tot}^{(at)}$ = 24.45$\cdot\ 10^{-4}~\AA^2$. A mean transverse energy gain $<\Delta E_\bot/\Delta z>_{at}$ = $8/a^3 \sigma_{tot}^{(at)} pv/2 <\theta_x^2>_{at}$ = 0.709 eV/$\mu$m is obtained for a somehow arbitrarily chosen cut-off angle of 0.02 rad. The mean number of collisions is $8/a^3 \sigma_{tot}^{(at)}$ = 4.31/$\mu$m.

\section{Electron-electron interactions}\label{interaction with electrons}
\subsection{General remarks}\label{general remarks}
The mean transverse energy gain $\overline{\Delta E_\bot^{el}}$ due to the scattering of the beam particle at atomic electrons in an interval $\Delta z=\Delta t~v$ can approximately be related to the mean electronic energy loss by the relation \cite{BelT81}
\begin{equation}
\frac{\overline{\Delta E_\bot^{(el)}}}{\Delta z}=\frac{\alpha_P}{2\gamma}\frac{\overline{\Delta E^{(el)}}}{\Delta z}
\label{TransverseEnergyGain}
\end{equation}
with $\alpha_P~\simeq ~0.5$.
The mean energy loss $\overline{\Delta E^{(el)}/\Delta z}$ can either be taken from tables, e.g. \cite{website:estar}, or calculated for higher energies than 1 GeV with the equation \cite{BelT81}
\begin{equation}
\frac{\overline{\Delta E^{(el)}}}{\Delta z} = \frac{2\pi (\hbar c \alpha)^2} {m_e c^2 \beta^2}\frac{8 Z}{a^3}L_e,
\label{electronMeanTransvEnergy}
\end{equation}
with the Sternheimer correction factor \cite[Eq. (52), Eq. (2), Table II]{SteP71}
\begin{eqnarray}\label{SternheimerCorrection}
L_e=\ln\Big(\frac{\gamma^2 2 m_e c^2 pv/2}{I^2}\Big)-2\beta^2-4.606\cdot\lg(\gamma)+4.636.
\end{eqnarray}
The quantity $I$ is the ionization potential for which $I$ = 89.4 eV has been used \cite{website:nist}, $\alpha$ the fine-structure constant, and $\beta~=~v/c$. With $\overline{\Delta E^{(el)}/\Delta z}$ = 737.82 eV/$\mu$m, from Eq. (\ref{electronMeanTransvEnergy}), one obtains from Eq. (\ref{TransverseEnergyGain}) $\overline{\Delta E_\bot^{(el)}/\Delta z}$ = 0.110 eV/$\mu$m.

There are different possibilities to proceed. The simplest one is to neglect scattering on electrons, leading to an underestimate of the transverse energy transfer. Another at the first glance also simple one would be to treat the electrons as a free gas and apply M{\o}ller scattering. However, a proper low energy cutoff is required since the M{\o}ller cross-section diverges at zero energy transfer. The third possibility applied here requires model assumptions on the generalized oscillator strength of the atom, as described below and in more detail in appendix \ref{appendix A}. In the latter, also the implications of the free electron gas approach are discussed in more detail.

\subsection{Scattering distribution for electrons}\label{scattering with electrons}
From Eq. (\ref{twodimcrossectionomega}) of appendix \ref{appendix A} one obtains for the angular differential cross-section
\begin{eqnarray}\label{electronicAngularDifferentialCrossection}
\frac{d\sigma^{(el)}}{d\Omega}(\theta) &=& \int\limits_{W_{min}(\theta)}^{W_{m}}dW\frac{d^2 \sigma^{(el)}}{dWd\Omega}(\theta,W)~~\Theta(\theta_{m}-\theta)+
\nonumber\\
& &\hspace{1.8 cm}
\frac{d\sigma_{Mo}}{d\Omega}(\theta)~~\Theta(\theta-\theta_{m}).
\end{eqnarray}
with the matching energy $W_{m}$ = 27~keV, and the matching angle $\theta_{m} = \theta_{max}$($W_m$) = 194.2 $\mu$rad from righthand side of Eq. (\ref{inequlitiesTwodimCrossSectionThetaWTheta}).
For $W > W_{m}$ all electrons of the carbon atom are regarded as quasi-free, and the M{\o}ller cross-section in the laboratory system, Eqs. (1) and (2), of Ref. \cite{Dal61} can be applied (read in Eq. (1) for the prefactor of last term $(\gamma-1)^2/\gamma^2$ instead of $(\gamma-1)^2/\gamma$). It should be mentioned, that at $W_{m}$ the cross-sections, first and second term of Eq. (\ref{electronicAngularDifferentialCrossection}), match with an accuracy of 1.9 \%. This is a remarkable result in view of the fact that both cross-sections were calculated by complete different approaches.

The total cross-section is the integral
\begin{eqnarray}\label{totalElectronicCrossection}
\sigma_{tot}^{(el)} = \int\limits_{\vartheta=0}^{\Theta_{max}}
d\vartheta~2\pi\sin\vartheta \frac{d\sigma^{(el)}}{d\Omega}(\vartheta).
\end{eqnarray}
For $\Theta_{max} = \arcsin \sqrt{2/(\gamma+3)}$= 0.03454 rad has been chosen, corresponding to the maximum energy loss of $(\gamma-1)m_e c^2/2$ = 427.5 MeV. The total cross-section for scattering at a single electron is $\sigma_{tot}^{(el)}$ = 5.576 $\cdot 10^{-4}~\AA^2$. This result agrees rather precisely with the integral $\int (d\sigma^{(el)}/dW)~dW = 5.577 \cdot 10^{-4}~\AA^2$ of Eq. (\ref{energydifferentialcrossSection}) in appendix \ref{appendix A}.

Again, at planar channeling we are interested in the differential scattering cross-section with respect to the $x$ coordinate. It is obtained by integration of Eq. (\ref{electronicAngularDifferentialCrossection}) over the $\vartheta_y$ coordinate and reads
\begin{eqnarray}\label{electronicAngularDifferentialCrossectionX}
\frac{d\sigma^{(el)}}{d\vartheta_x}(\vartheta_x) = \int\limits_{\vartheta_y=-\Theta_{y}}^{\Theta_{y}}
d\vartheta_y \frac{d\sigma^{(el)}}{d\Omega}\Big(\sqrt{\vartheta_x^2+\vartheta_y^2}\Big)
\end{eqnarray}
with $\Theta_{y}=\sqrt{\Theta_{max}^2-\vartheta_x^2}$.
The scattering distribution is the normalized angular distribution of Eq. (\ref{electronicAngularDifferentialCrossectionX})
\begin{equation}
P^{(el)}(\theta_x)=\frac{d\sigma^{(el)}}{d\theta_x}(\theta_x)/\sigma_{tot}^{(el)}
\label{probabilityDistributionEl}
\end{equation}
which is depicted also in Fig. \ref{ProbabilityWnorm}, dot-dashed line. It can be approximated by the heuristic function\footnote{The parameters are  $A_{855}$ = 3.5 $\cdot 10^{-12}$, $a_{855} =7.0\cdot 10^{-19} $,  $b_{855} =3.6 \cdot10^{-11} $ and $c_{855} =7.0 \cdot10^{-8} $. The approximations are better than about  $\pm$ 10 \%  in the interval  0.1 $\mu$rad  $< \lvert\theta_x \rvert<$ 20 mrad. Below the lower and above the upper limit the approximations become significantly worse.}
\begin{eqnarray}\label{Pel855MeVapprox}
P^{(el)}(\theta_x)=\frac{A}{\lvert\theta_x^3 \rvert + c \cdot \theta_x ^2 + b\cdot \lvert\theta_x\rvert + a}.
\end{eqnarray}

The mean squared scattering angle
\begin{equation}
<\theta_x^2>_{el}~ = \int\limits_{-\Theta_x}^{\Theta_x} \theta_x^2 \cdot P^{(el)}(\theta_x)d\theta_x
\label{meansquaredScatteringAngle}
\end{equation}
is a function of the integration limit $\Theta_x$. For $\Theta_{max}$ = 0.03454 rad $<\theta_x^2>_{el} = 5.370 \cdot 10^{-11}$ results. For the calculation of the mean transverse energy gain, which is of relevance for channeling, one obtains $<\Delta E_\bot/\Delta z>_{el}$ = $(8~Z/a^3)$ $\sigma_{tot}^{(el)}pv/2 <\theta_x^2>_{el}$ = 0.136 eV/$\mu$m. This mean transverse energy gain for electron-electron collisions is a factor of 5.2 less than the corresponding number for electron-atom collisions. The mean number of collisions per atom and unit longitudinal length interval is $(8Z/a^3)\cdot\sigma_{tot}^{(el)}$ = 5.899/$\mu$m,

The electron density $n_{el}(x)$ across the channel is a function of the coordinate $x$. It has been derived from the potential $u(x)$ with the aid of the one dimensional Poisson equation. The result is shown in Fig. \ref{potential} (b). The mean number of collisions at the lateral position $x_k$ in an interval $\Delta z_k=\Delta t_k~c$ is
\begin{equation}
\Delta n_k=n_{el}(x_k)~\sigma_{tot}^{(el)}~\Delta z_k.
\label{meanElectronicCollisions}
\end{equation}
For this equation it has been assumed that the number of collisions depends only on the electron density, i.e., a possible dependence on the transverse coordinate $x$ of the cross-section has been neglected by the replacement with its mean $\sigma_{tot}^{(el)}$. However, considering a certain fraction of the electrons as quasi-free "sea" electrons, say 20 \%,  for which the cross section can be described by M{\o}ller scattering with a low energy cutoff of 6 eV, does not change the final simulation results significantly. The probability distribution for $n$ collisions at a mean $\Delta n_k$ is given by the Poisson-distribution function
\begin{equation}
P_n(k)=\frac{(\Delta n_k)^n}{n!}\exp(-\Delta n_k).
\label{PoissonDistribution}
\end{equation}
In the simulation $N_k = \mbox{Random}_{n}[P_{n}(k)]$ is randomly generated. For $N\geq 1$ the change of the scattering angle is
\begin{equation}
\Delta\vartheta_{k}^{(el)}= \sum_{i=1}^{N_k}\mbox{Random}_{i,\theta_x}[P^{(el)}(\theta_x)].
\label{electronicMeanTransvAngle}
\end{equation}

\section{Transverse energy transfer at collisions}\label{transverse energy tranfer}
The Monte Carlo simulated change of the scattering angle is with Eqns. (\ref{changeScatteringAngle}) and (\ref{electronicMeanTransvAngle}) $\Delta \vartheta_k=\Delta\vartheta_{k}^{(at)} +\Delta\vartheta_{k}^{(el)}$, and the new scattering angle at the step from $x_k\rightarrow x_{k+1}=x_k+\Delta x_k$ is given by
\begin{equation}
\vartheta_{k+1} =\vartheta_k+\Delta\vartheta_{k}= \pm\sqrt{2\big(E_{\bot,k}-U(x_{k})\big)/pv}+\Delta\vartheta_{k}
\label{newScatteringAngle}
\end{equation}
with $U(x_{k})$ the potential energy at position $x_k$.
A scattering by the angle $\Delta\vartheta_{k}$ results in the new transverse energy
\begin{eqnarray}\label{NewEperp}
E_{\bot,k+1}& = &E_{\bot,k}+
\nonumber\\
& &\hspace{-0,7 cm}
pv/2\cdot(\Delta\vartheta_k)^2 \pm\sqrt{2pv\cdot\big(E_{\bot,k}-U(x_{k})\big)}~~\Delta \vartheta_k.
\end{eqnarray}
The plus sign in front of the square root holds for $\Delta x_k>0$, or if $E_{\bot,k+1}-U(x_{k+1})$ is negative, the minus sign for $\Delta x_k<0$. How these equations come about is explained in Appendix \ref{appendix B}.
Since $\Delta \vartheta_k$ has both signs which are distributed with equal probability, the first term in the second row of  Eq. (\ref{NewEperp}) $\Delta E_{\perp,k}^{(\text{drift})}$ results in a drift while the second one $\Delta E_{\perp,k}^{(\text{diff})}$ in fluctuations of $E_{\bot}$. Therefore, both terms attribute to de-channeling while re-channeling is effected only by the second term.

With these definitions instantaneous drift and diffusion coefficients may be defined as
\begin{equation}
\tilde{d}_1(E_{\perp,k},~x_k)=\Delta E_{\perp,k}^{(\text{drift})}/\Delta z_k=p v/2 \cdot (\Delta\vartheta_k)^2/\Delta z_k,
\label{differentialDriftCoeff}
\end{equation}
and
\begin{eqnarray}\label{differentialDiffusionCoeff}
\tilde{d}_2(E_{\perp,k},~x_k) &=& 1/2 \cdot (\Delta E_{\perp,k}^{(\text{diff})})^2/\Delta z_k =
\nonumber\\
& &\hspace{-1,4 cm}
=1/2\cdot 2 p v \big(E_{\bot,k}-U(x_{k})\big) \cdot (\Delta\vartheta_k)^2/\Delta z_k.
\end{eqnarray}
In contrast to the coefficients $D_1(E_\perp)$ and $D_2(E_\perp)$ which enter the Fokker-Planck equation as mean values with respect of one oscillation period for a given $E_\perp$, these instantaneous quantities are functions of both, the variables $E_\perp$ and $x$. In addition, they have a functional dependence on the penetration depth $z$. We come back to this fact in chapter \ref{sectionFokkerPlanck}.

\section{Preparatory calculations}\label{calculations}
\subsection{Test of the simulation model for "amorphous diamond"}
\begin{figure}[tbh]
\centering
    \includegraphics[angle=0,scale=0.45,clip]{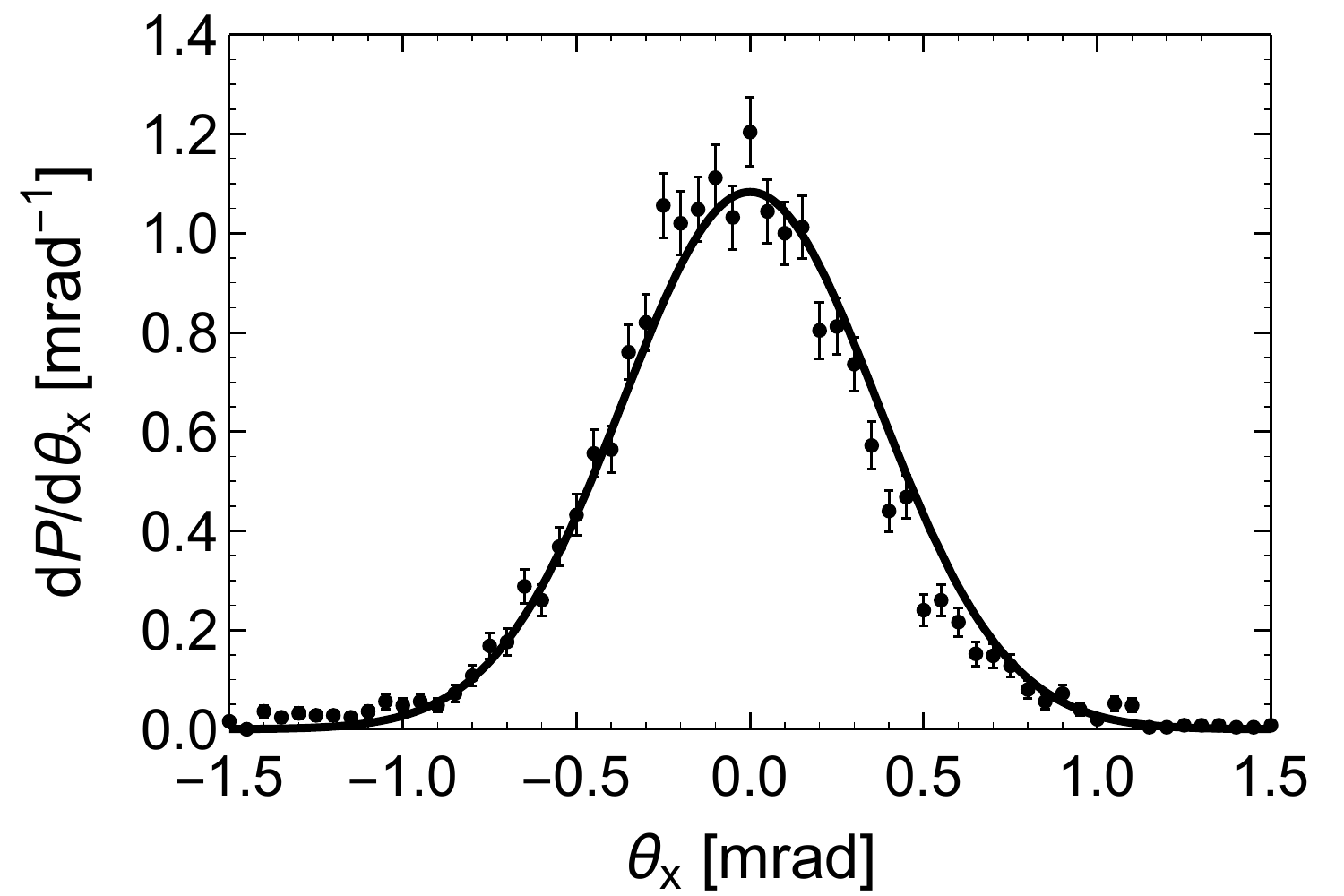}
\caption[]{Simulation of the scattering distribution at the passage of 855 MeV electrons through an amorphous carbon foil with the density of diamond and a thickness of 120 $\mu$m. A number of 5,000 tracks were simulated. The full curve represents the projected Gaussian angular distribution according to the Particle Data Group with $\theta_{plane}^{rms}=0.368$ mrad, which is quite similar to that one published by G.R. Lynch and O.I. Dahl \cite{LynD91} with 98\% in the sample and $\theta_{plane}^{rms}=0.361$ mrad.} \label{Simulation50muDiam}
\end{figure}

The formalism described above has been checked by a simulation of scattering of 855 MeV electrons passing an amorphous carbon foil with the density of diamond and a thickness of 120 $\mu$m. The mean number of collisions amounted to 517.6 and 658.3 for collisions with the atoms and electrons, respectively. The quantities $\Delta\vartheta_{k}^{(at)}$ and $\Delta\vartheta_{k}^{(el)}$ were calculated with Eqns. (\ref{changeScatteringAngle}) and (\ref{electronicMeanTransvAngle}), respectively. The scattering distribution as function of $\theta_x=\Delta\vartheta_{k}^{(at)}+\Delta\vartheta_{k}^{(el)}$ is shown in Fig. \ref{Simulation50muDiam}. It should be mentioned that the width of the electronic distribution amounts to about 35 \% of the width of the atomic distribution which can not be neglected. The total distribution exhibits a slight asymmetry, clearly recognizable when compared with the also shown Gaussian angular distribution, which is typical for a Poisson distribution. In conclusion, the overall agreement is quite good. This fact provides some confidence in trusting the simulation calculations for the de-channeling length to be described in section \ref{results}.
\subsection{Initial occupation probability}
\begin{figure}[tbh]
\centering
    \includegraphics[angle=0,scale=0.45,clip]{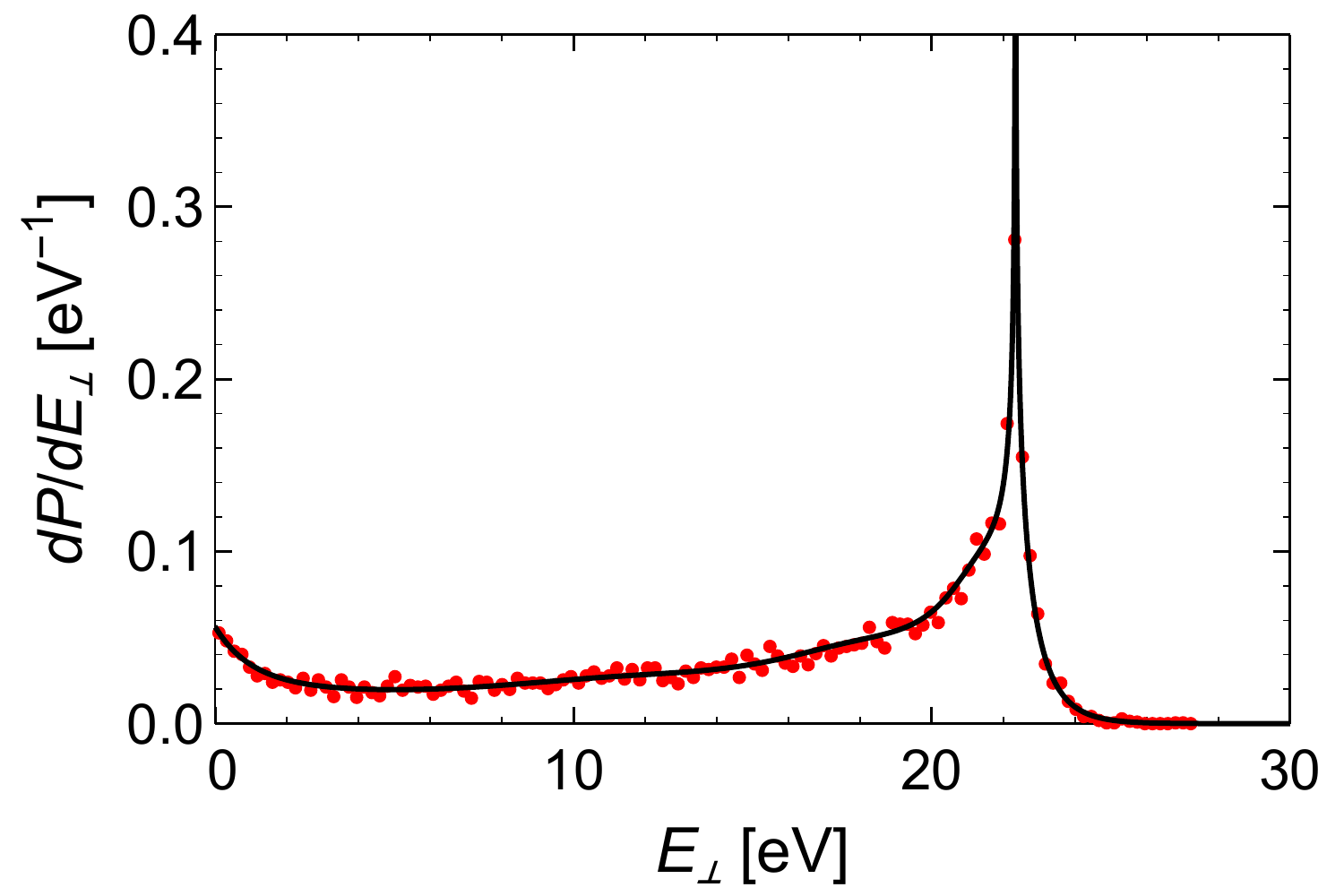}
\caption[]{Initial distribution for a straight diamond crystal. The black full curve represents the analytical expression of Eq. (15) of Backe et al. \cite{BacL08}, the red dots are simulation calculation with 10,000 trials.} \label{InitialCalcSim}
\end{figure}

For the initial occupation probability in the potential pocket a uniform distribution of the electron density across the transverse $x$ coordinate, and a Gaussian scattering distribution with standard deviation $\sigma '_{x}$ = 12.7 $\mu$rad for the angular divergence were assumed. The initial occupation probability can be analytically calculated according to Backe et al. \cite[Eq. (15)]{BacL08}. The initial distribution is shown in Fig. \ref{InitialCalcSim} for a straight diamond crystal. Part of the distribution is located in the continuum, i.e., about 12 \%  occupy states with $u>u_0$ = 23.96 eV.

\section{Results}\label{results}
\subsection{Results for channeling at (110) planes of diamond and discussion}\label{results (110) planes}
\label{channelingPlane}
Now all ingredients are available to investigate the de-channeling process. Simulations have been performed for a target thickness of 200~$\mu$m for three consecutive categories, primary channeling mode, de-channeling mode, and secondary channeling mode. The channeling history is terminated either if the electron leaves the potential boundaries in the secondary channeling mode, if it will not be de-channeled after a distance of 100~$\mu$m, or if it reaches the maximum transverse energy of 10 $u_0$ = 240 eV.  Typical examples of trajectories are shown in Appendix \ref{appendix C}.
\begin{figure}[tbh]
\centering
    \includegraphics[angle=0,scale=0.55,clip]{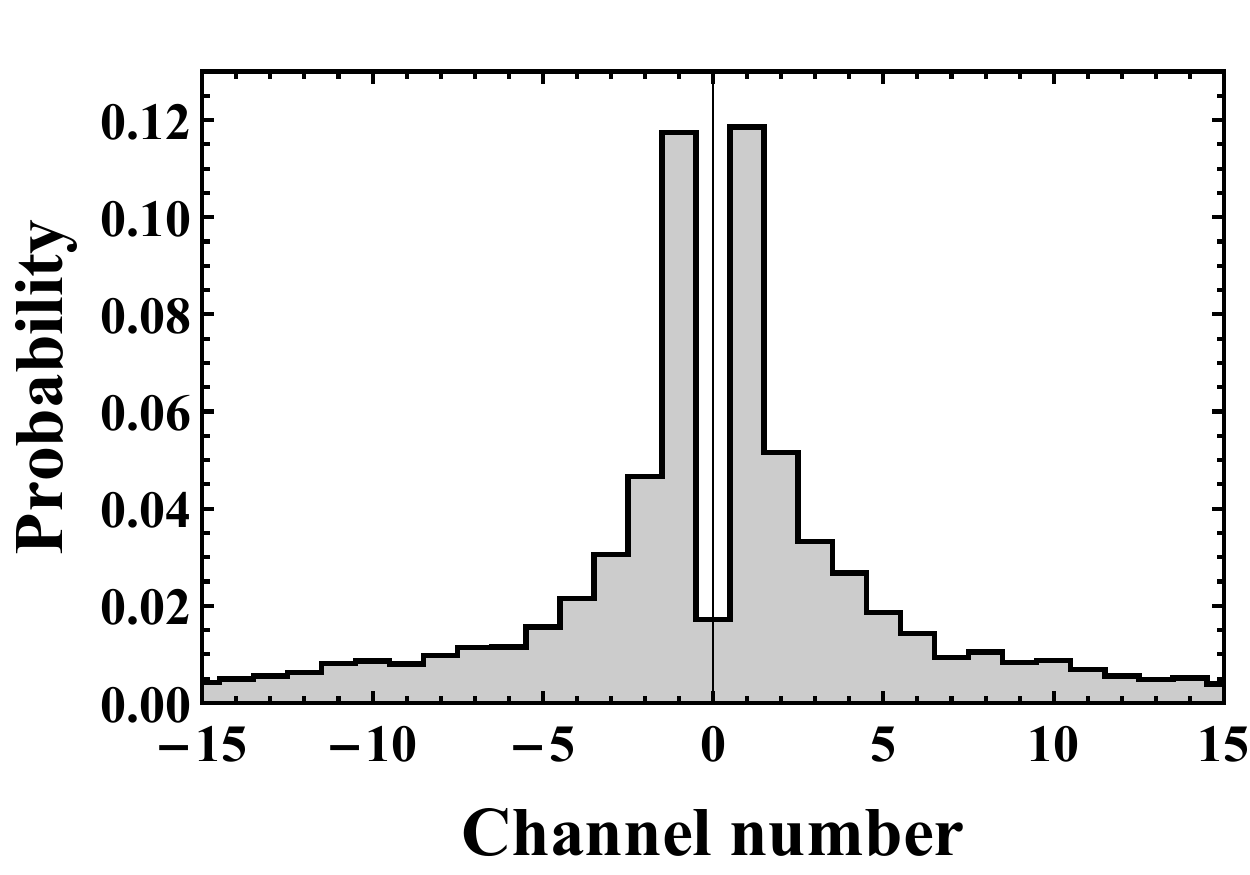}
\caption[]{Normalized channel number distribution. Channel 0 is the primary channel in which the particle was originally captured, neighbouring ones in which the particle re-channelled are counted from there.} \label{ChannelingNumberDistribution}
\end{figure}

Fig. \ref{ChannelingNumberDistribution} shows the channel number distribution for re-channeling. It is interesting to notice that the largest probability for a re-channeling occurs at the direct neighboring channels. This is expected since de- and re-channeling are both governed by the same scattering distribution shown in Fig. \ref{ProbabilityWnorm} which is largest at small energy transfers. The larger the transverse energy, the smaller the probability for a re-channeling event. The electron may re-channel to channel 0 if it returns after de-channeling, i.e. crossing the boundary  $x = \pm d_p/2$, again to channel 0. The integrated probability for re-channeling up to the second neighboring channels amounts to 33 \%.
\begin{figure}[b]
\centering
    \includegraphics[angle=0,scale=0.65,clip]{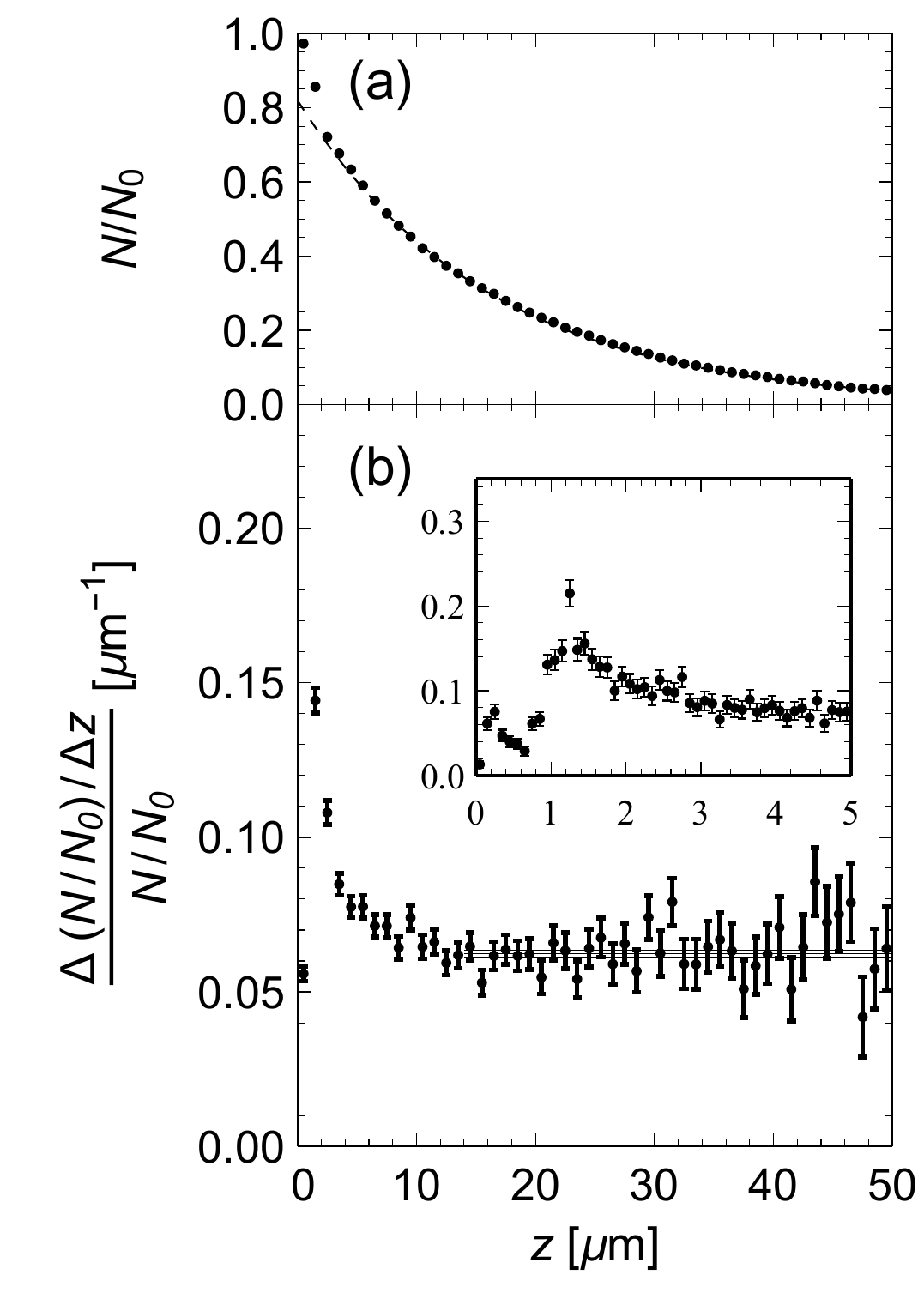}
\caption[]{(a) Primary channel occupation and (b) approximation of the de-channeling rate as function of the penetration depth $z$, with details shown in the inset. The channel occupation is normalized to unity with $N_0$ = 10,000 events. The horizontal error band indicates the mean value $\overline{\lambda(z)}$ = (0.0624 $\pm$ 0.0011)/$\mu$m as obtained for $14 \leq z/\mu m \leq 100$.} \label{dechannelinRateFirst}
\end{figure}
\begin{figure}[b]
\centering
    \includegraphics[angle=0,scale=0.65,clip]{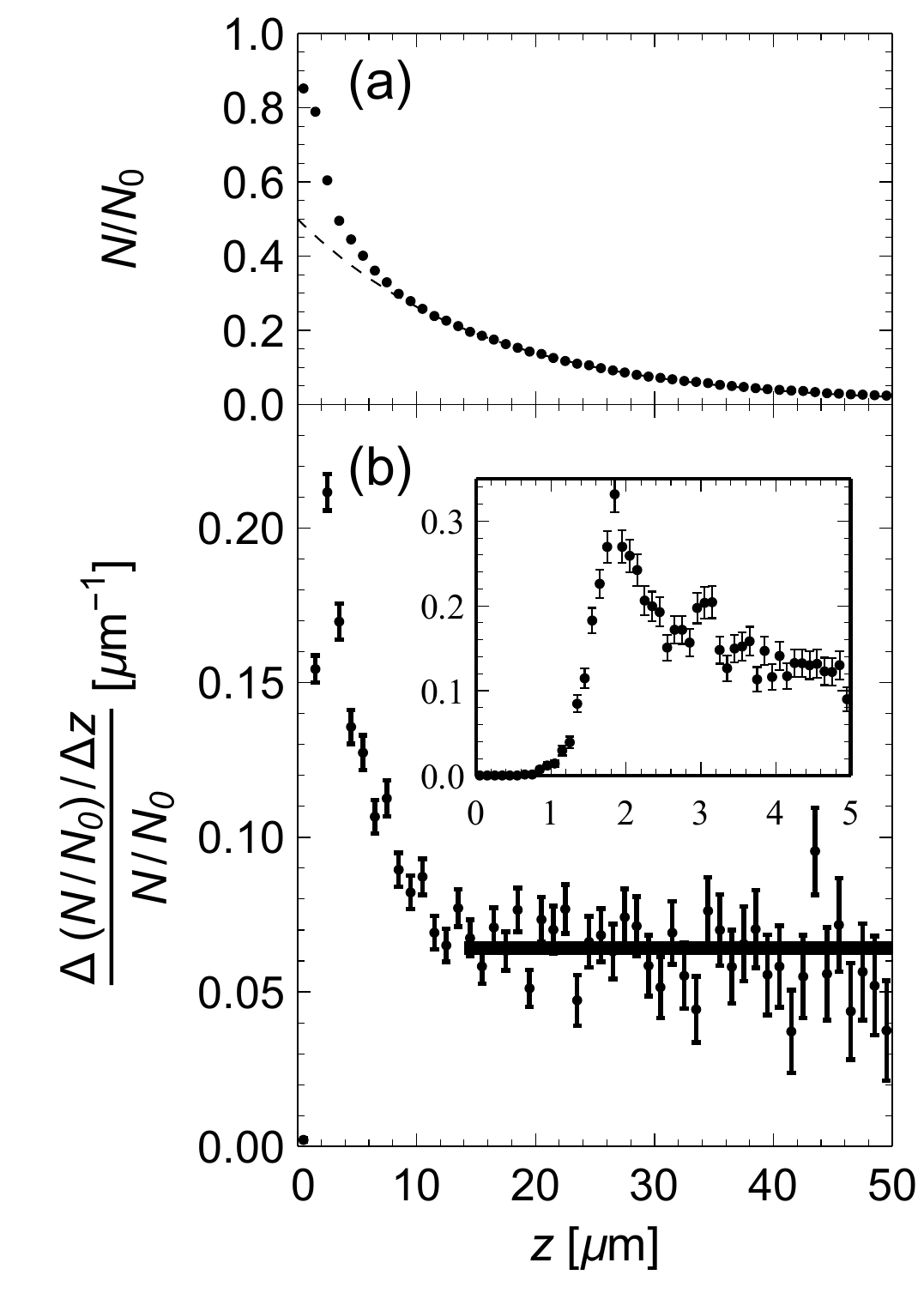}
\caption[]{Same  as Fig. \ref{dechannelinRateFirst} for re-channeled particles. The horizontal error band indicates the mean value $\overline{\lambda(z)}$ = (0.0641 $\pm$ 0.0014)/$\mu$m as obtained for $14 \leq z/\mu m \leq 100$.\\ \\} \label{dechannelinRateSecond}
\end{figure}

Fig. \ref{dechannelinRateFirst} shows in panel (b) an approximation of the instantaneous transition rate
\begin{eqnarray}
\lambda _{de}(z) = -\frac{f_{de}'(z)}{f_{de}(z)} =
\lim_{\substack {\Delta z\to 0 \\ N \to \infty}} \frac{\Delta {(N}(z)/N_0)/\Delta z }{N(z)/N_0}
\label{dechRate}
\end{eqnarray}
as derived from the simulated channel occupation numbers \newline $\Delta {(N}(z)/N_0)/\Delta z$ and the channel occupation $N(z)/N_0$ depicted in panel (a). For $z$ less than about 14 $\mu$m the instantaneous transition rate is larger than the constant mean value for $z>$ 14 $\mu$m, see panel (b). This is a consequence of the fact that the captured electrons need a certain relaxation depth to achieve statistical equilibrium. The structure of the rate for $z\leq5 \mu$m is shown in the inset. The constant mean value at statistical equilibrium is $\overline{\lambda_{de,1}^{plane}(z)}$ = (0.0624 $\pm$ 0.0011)/$\mu$m, as calculated for the interval $14 \leq z/\mu m \leq 100$. A fraction of 18\% of particles de-channel in the mean within the relaxation depth.

A de-channeled particle may re-channel. Here such an process is defined by the requirement that the particle must be reflected from the potential wall at least once. Results for these secondary channeling process are shown in Fig. \ref{dechannelinRateSecond}. The rate exhibits similar features as shown for the primary process. The equilibrium depth turns out to be $\overline{\lambda_{de,2}^{plane}(z)}$ = (0.0641 $\pm$ 0.0014)/$\mu$m for $14 \leq z/\mu m \leq 100$ which is in accord with the primary de-channeling rate. However, a much larger fraction of 50\% is lost within the relaxation depth $z<14\mu$m.

Alternatively, an effective de-channeling length $\Lambda_{de}$ may be defined via the integral $\int_{z_0}^{\Lambda_{de}} \lambda_{de}(z)~dz$ = 1 with $z_0$ the entrance position into the channel. One obtains from Fig. \ref{dechannelinRateFirst} (b) and \ref{dechannelinRateSecond} (b) for $\Lambda_{de} \approx$ 13 and $\approx$ 8 $\mu$m for the first and second de-channeling length, respectively. These significant different values may be explained by the fact that re-channeling happens mainly in the middle of the channel where the overlap with the atomic density is largest. Therefore, a shorter equilibration time is needed in comparison with the primary channel occupation probability which is distributed over the whole channel width.

\subsection{Channeling in bent crystals}
\label{channelingBent}
For production of undulator-like radiation the crystal must be bent. In the following, the diamond crystal is assumed to have dimensions like the silicon crystal described in Ref. \cite{MazB14}, i.e., thickness of 30.5 $\mu$m and bending radius of 33.5 mm. Calculations were performed as described in section \ref{channelingPlane}. A primary de-channeling rate $\overline{\lambda_{de,1}^{bent}(z)}$ = (0.0668 $\pm$ 0.0010)/$\mu$m has been obtained for the interval $14 \leq z/\mu m \leq 100$ for 75 \% of particles imping the crystal. For re-channeled particles the de-channeling rate turned out to be  $\overline{\lambda_{de,2}^{bent}(z)}$ = (0.0685 $\pm$ 0.0025)/$\mu$m, for the same interval with an efficiency of only 14 \% of particles imping the crystal. While the primary de-channeling rate for the bent crystal increases by only 7 \% in comparison with the plane one, the efficiency decreases dramatically.

\begin{figure}[h]
\centering
    \includegraphics[angle=0,scale=0.45,clip]{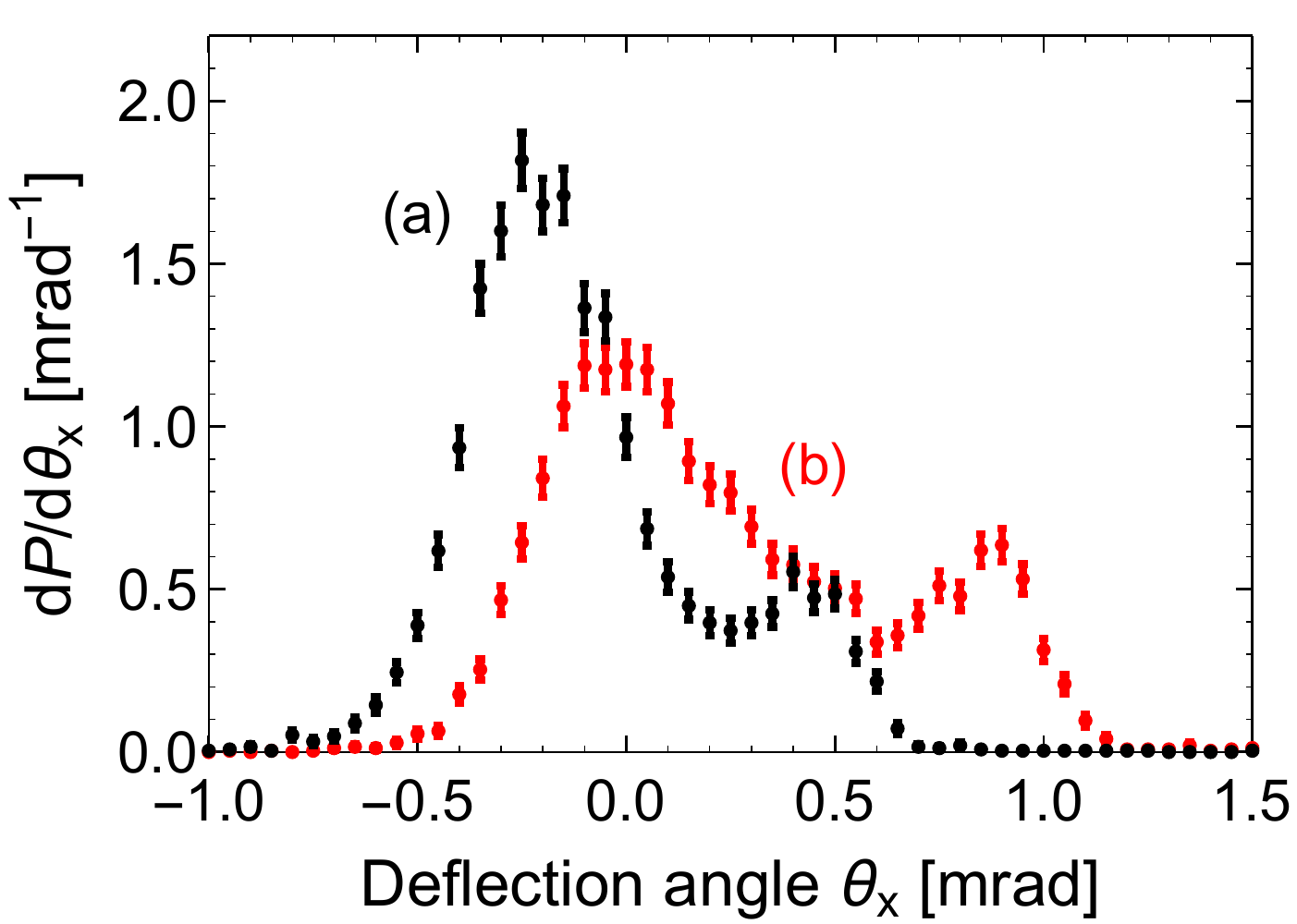}
\caption[]{Beam profiles for a 30.5 $\mu$m thick diamond crystal with a bending radius of 33.5 mm for (a) an entrance angle $\psi_0$ = -0.45 mrad, and (b) $\psi_0$ = 0 mrad.} \label{beamProfilesBent}
\end{figure}

In addition, the beam profile at the exit of the crystal has been calculated. The result is shown in Fig. \ref{beamProfilesBent}. The profiles exhibit striking similarities with the results shown by Mazzolari at al.  \cite[Fig. 3]{MazB14}. In particular the deflection peaks due to channeling in (a) and (b), as well as the volume reflection shift in opposite direction for the tilted crystal in (a) can clearly be recognized.

\subsection{Comparison with results from the MBN explorer package}
\begin{figure}[tb]
\centering
    \includegraphics[angle=0,scale=0.60,clip]{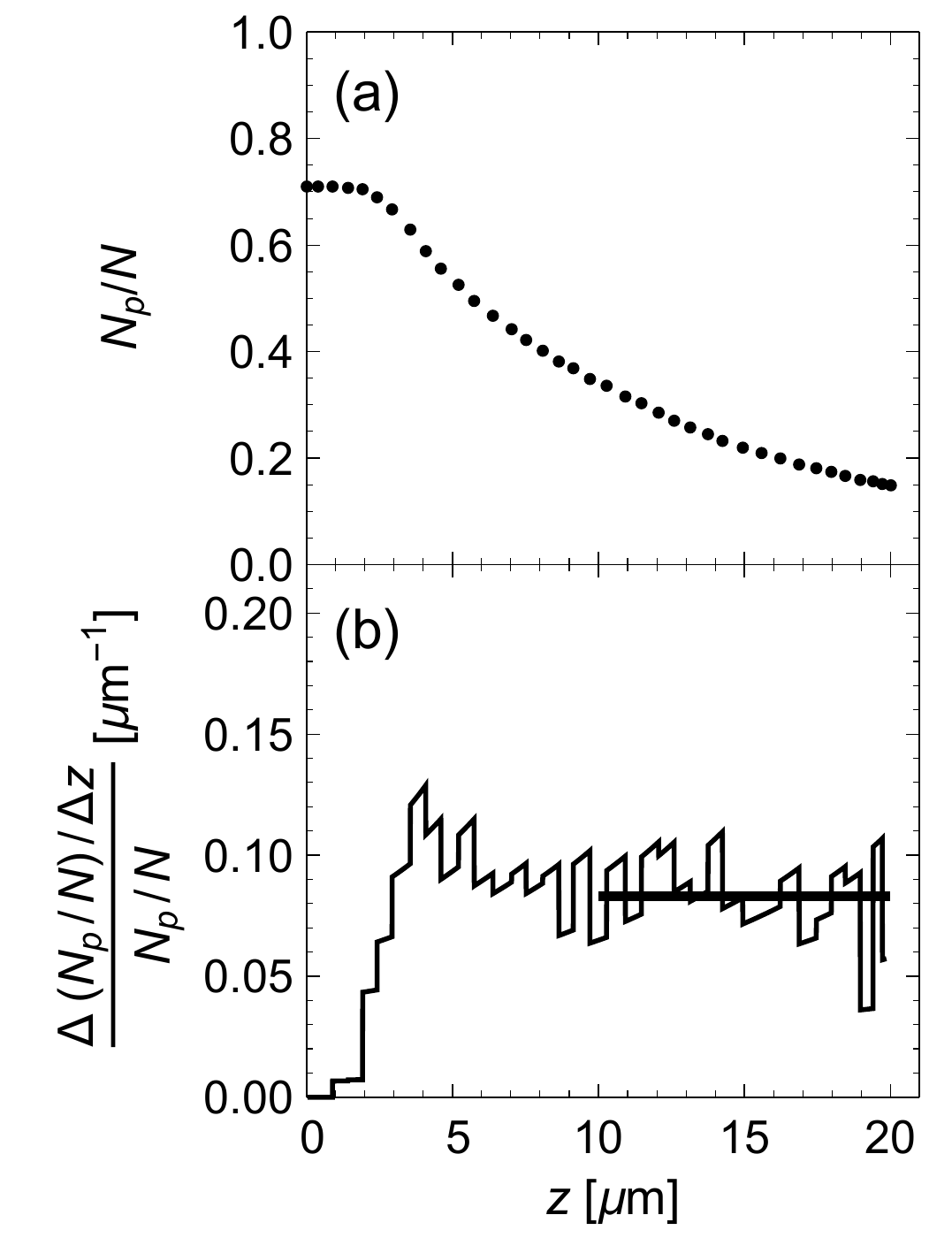}
\caption[]{(a) Digitized fraction $N_p/N$ of 855 MeV electrons in a straight diamond (110) crystal channel as taken from Pavlov et al. \cite[Fig. 5 (c)]{PavK20}, red curve, and (b) differential quotient normalized to the fraction of electrons in the (110) channel. The estimated equilibrium de-channeling rate in the shown interval $\overline{\lambda(z)}$ = 0.083/$\mu$m has an error of about $\pm$0.01/$\mu$m, mainly due to digitization. } \label{AnalysisFig5c}
\end{figure}

Simulation calculations have been performed on the basis of the rather sophisticated MBN Explorer package as described in a number of publications of the group around A.V. Korol, A.V. Solov'yov et al. In the paper of A.V. Pavlov et al. \cite{PavK20} simulation calculations were done also for channeling of 855 MeV electrons in diamond single crystals. Unfortunately, it is not possible to compare the results directly with results obtained in this paper. Therefore, Fig. 5 (c) (red curve) of Ref. \cite{PavK20} was digitized. The result is depicted in Fig. \ref{AnalysisFig5c} (a). In Fig. \ref{AnalysisFig5c} (b) an approximation of $\lambda_{p}(z)$ according to Eq. (\ref{dechRate}) is shown.
Particularly striking are the zero-valued de-channeling transition rates for small penetration depths. This behaviour reflects the acceptance criterion, meaning that a particle is accepted to be in the channeling mode only if it changes sign of its transverse velocity twice. The mean equilibrium de-channeling length obtained in the interval between $10\leq z/\mu \text{m} \leq 20$ amounts to $\lambda _{p}$ = 0.083/$\mu$m, and is about (33$\pm$16)\% larger than that obtained in this work. Whether this is a significant difference, or not, is in view of the rather large uncertainty currently not clear.

\section{Impact on the Fokker-Planck equation}\label{sectionFokkerPlanck}
The solution of the Fokker-Planck equation for planar channeling experiments in straight crystals with electrons at the 855 MeV Mainz Microtron MAMI has been described in our previous work \cite{BacL08}. The drift coefficient, and in turn also the diffusion coefficient, were calculated with the Kitagava-Ohtsuki approximation \cite{KitO73} which in \cite{BacL08} was represented in the following form
\begin{eqnarray}\label{driftD1Kitagava}
D_1 (E_\bot) &=& \frac{E_s^2}{2 pv X_0}\times
\nonumber\\
& &\hspace{0 cm}
\int_{x_{min}}^{x_{max}}\frac{dP}{dx}(E_{\bot},x)\frac{d_p}{\sqrt{2\pi}u_1} \exp(-x^2/2u_1^2)~dx ~~~
\end{eqnarray}
with
\begin{eqnarray}\label{ProbabilityDistribution}
\frac{dP}{dx}(E_{\perp},x)=\frac{2}{T(E_{\perp})\cdot v}\sqrt{\frac{\gamma\cdot m_e c^2}{{2\big(E_{\perp}}-u(x)\big)}}~~,
\end{eqnarray}
\begin{eqnarray}\label{Period}
T(E_{\perp})\cdot v = 2\int\limits_{x_{min}}^{x_{max}} \sqrt{\frac{\gamma \cdot m_e c^2}{2\big(E_{\perp}-u(x)\big)}}~~dx.
\end{eqnarray}
The integration limits $x_{min}$ and $x_{max}$ are roots of the equation $E_\bot=u(x)$. The quantity $X_0$ = 0.1213 $\cdot 10^6~\mu$m is the radiation length \cite{Pat17}. The results of the solution of the Fokker-Planck equation depend strongly on the choice of the parameter $E_s$ since it enters quadratically. In our previous work $E_s$ = 10.6 MeV was chosen \cite{BacL18}, resulting in a de-channeling length of 41.04 $\mu$m. This value deviates by the large factor of about 2.6 from the simulation calculation result  of 1/$\overline{\lambda(z)}$ = 15.8 $\mu$m quoted in the previous section \ref{results (110) planes}. In the following the diffusion coefficients entering the Fokker-Planck equation will be reanalyzed for both, (i) scattering at atoms only, and (ii) including scattering at electrons with the formalism described above.

(i) For scattering at atoms drift and diffusion coefficients may be written as
\begin{eqnarray}\label{driftD1atomic}
D_1^{(at)} (E_\bot) &=& \frac{pv}{2}<\theta_x^2>_{at}\frac{8}{a^3}~\cdot\sigma_{tot}^{(at)}\times
\nonumber\\
& &\hspace{-1 cm}
\int\limits_{x_{min}}^{x_{max}} \frac{dP}{dx}(E_{\bot},x) \frac{d_p}{\sqrt{2\pi}u_1} \exp(-x^2/2u_1^2) dx
\end{eqnarray}
and
\begin{eqnarray}\label{driftD2atomic}
D_2^{(at)} (E_\bot)&=& \frac{pv}{2}<\theta_x^2>_{at}\frac{8}{a^3}~\cdot\sigma_{tot}^{(at)}\times
\nonumber\\
& &\hspace{-2 cm}
\int\limits_{x_{min}}^{x_{max}}  2 (E_\bot-u(x)) \frac{dP}{dx}(E_{\bot},x) \frac{d_p}{\sqrt{2\pi}u_1} \exp(-x^2/2u_1^2) dx.
\end{eqnarray}
Three terms in Eqs. (\ref{driftD1atomic}) and (\ref{driftD2atomic}) can be distinguished which have the following meanings: The pre-factor $(pv/2)<\theta_x^2>_{at}$ is the mean transverse energy gain at a single elementary scattering process, the pre-factor $(8/a^3)~\cdot\sigma_{tot}^{(at)}$ represents the mean number of collisions per unit path length, and the integrals represent averages for one oscillation period accounting for the distribution of the scattering centers across the channel. Defining in analogy to Eq. (\ref{driftD1Kitagava}) an $E_{s,m}^2 = (pv)^2 X_0$ $<\theta_x^2>_{at} \cdot (8/a^3) \cdot\sigma_{tot}^{(at)}$, the microscopically defined scattering parameter $E_{s,m}$ = 12.13 MeV is obtained. With this value the de-channeling length as derived from the solution of the Fokker-Planck equation reduces from $L_{de}^{FP}$ = 41.04 $\mu$m for $E_s$ = 10.6 MeV to the significantly lower value 31.3 $\mu$m.

\begin{figure}[H]
\centering
    \includegraphics[angle=0,scale=0.6,clip]{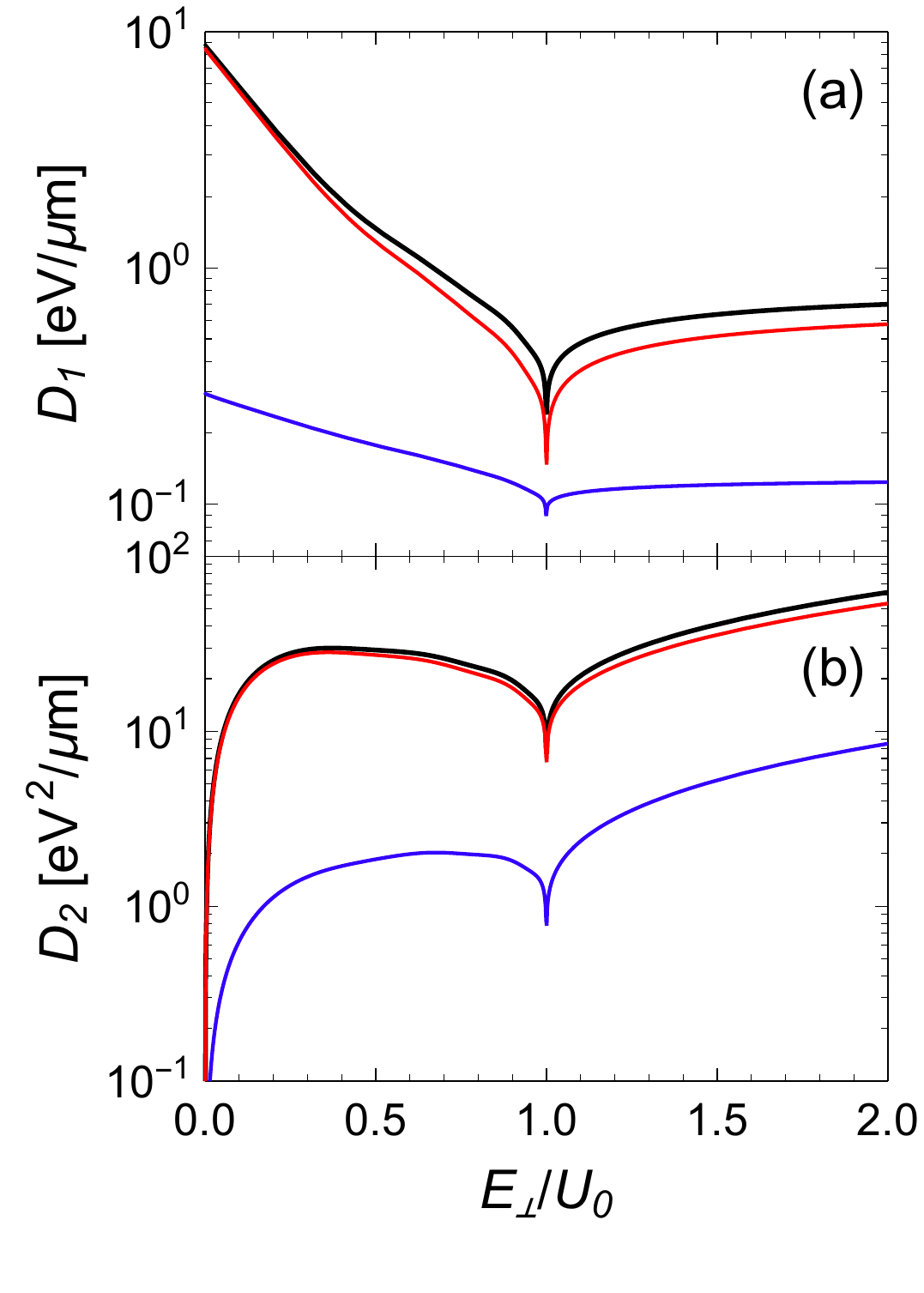}
\caption[]{(a) Drift coefficients $D_1$, and (b) diffusion coefficient $D_2$ as function of the normalized transverse energy $E_\perp/U_0$ for a straight diamond crystal with $U_0 = u_0$. Red curve: contributions from atoms, blue from electrons, and black is the total.} \label{DriftDiffusionAllLog}
\end{figure}

(ii) For scattering at electrons the diffusion coefficients are defined in analogy to the scattering at atoms as
\begin{eqnarray}\label{driftD1Electron}
D_1^{(el)} (E_\bot)&=&\frac{pv}{2}  <\theta_x^2>_{el}
\int\limits_{x_{min}}^{x_{max}} \frac{dP}{dx}(E_{\bot},x)~n_{el}(x)~\sigma_{tot}^{(el)} dx~~~~~
\end{eqnarray}
and
\begin{eqnarray}\label{driftD2Electron}
D_2^{(el)} (E_\bot)=\frac{pv}{2}  <\theta_x^2>_{el}\times
\nonumber\\
& &\hspace{-2,6 cm}
\int\limits_{x_{min}}^{x_{max}} 2 (E_\bot-U(x)) \frac{dP}{dx}(E_{\bot},x)~n_{el}(x)~\sigma_{tot}^{(el)} dx.~~~
\end{eqnarray}
The Fokker-Planck equation will be solved for the sum of the atomic and electronic contributions
\begin{equation}
D_1(E_\bot) = D_1^{(at)}(E_\bot)+D_1^{(el)}(E_\bot),
\label{driftD1total}
\end{equation}
\begin{equation}
D_2(E_\bot) = D_2^{(at)}(E_\bot)+D_2^{(el)}(E_\bot).
\label{driftD2total}
\end{equation}
Drift and diffusion coefficients are shown in Fig. \ref{DriftDiffusionAllLog}. Although the contribution of the electrons seems to be small, scattering at electrons has a significant impact, i.e., the de-channeling length reduces to 21.8 $\mu$m. This value is still 38 \% larger than that one obtained by simulation calculations.
\begin{figure}[tbh]
\centering
    \includegraphics[angle=0,scale=0.5,clip]{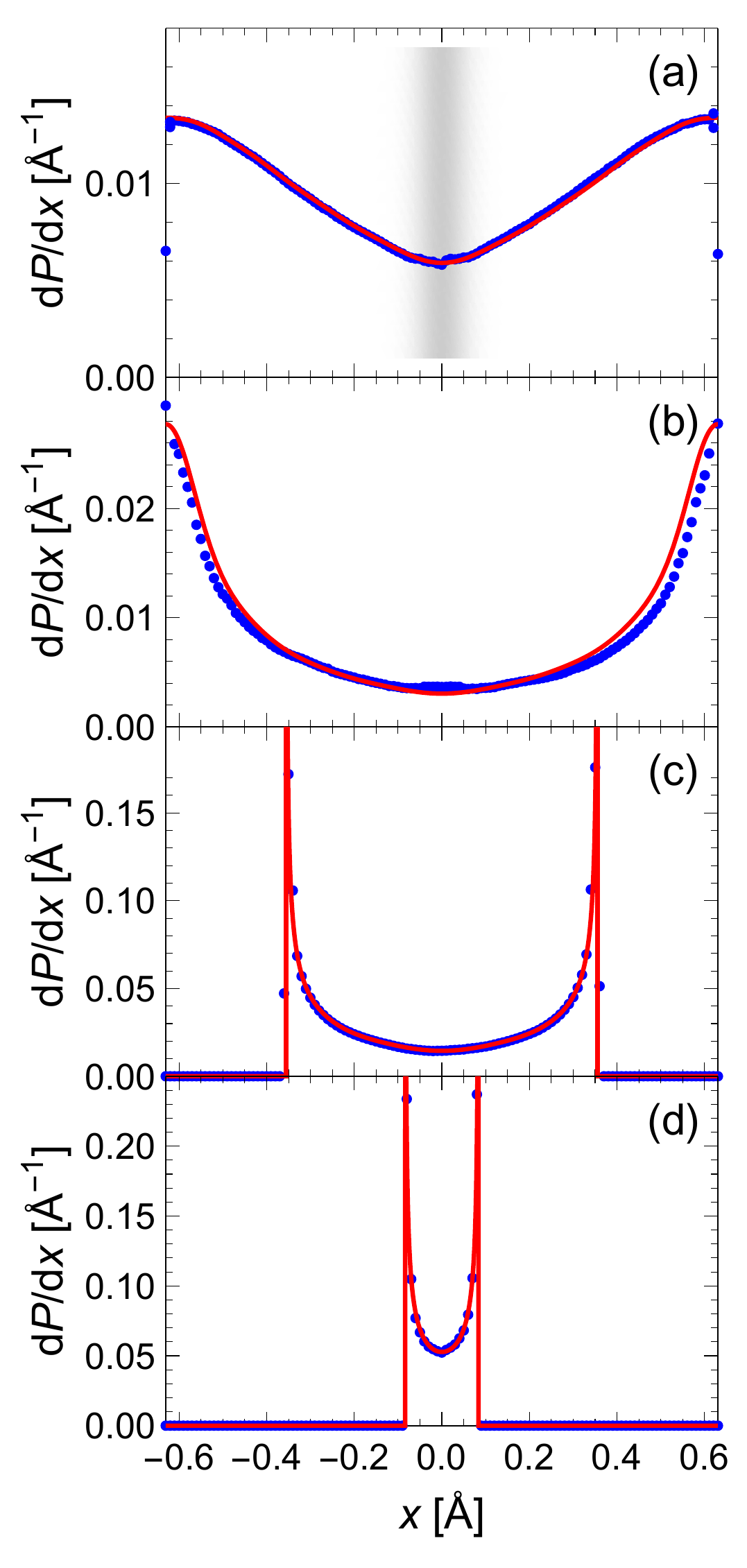}
\caption[]{Transverse probability distributions. Red full curve depict calculations with Eq. (\ref{ProbabilityDistribution}), blue dots simulations. (a) $\overline{E_\perp}$ = 29.8 eV well above the barrier,  (b) 24.2 eV at the barrier, (c) 19.7 eV below the barrier, and (d) 4.7 eV well below the barrier. The shadowed region indicates the nuclear density distribution.} \label{transverseProbDist}
\end{figure}

In order to examine possible reasons for the remaining difference between the simulation calculation and the solution of the Fokker-Planck equation, in Fig. \ref{transverseProbDist} simulation results for the probability distribution are compared with calculations of $dP/dx(E_\perp,x)$ with the aid of Eq. (\ref{ProbabilityDistribution}). Both probability distributions look very similar. Only a rather small enhancement of the simulation results at the overlap region with the atomic density distribution may be discernable in panel Fig. \ref{transverseProbDist} (b) for $\overline{E_\perp}$ = 24.2 eV. Tikhomirov discussed \cite[ch. 3.4]{Tik17} that the probability distribution Eq. (\ref{ProbabilityDistribution}) cannot be applied because of large transverse energy variations at domains where the atomic density is large. This statement can more or less be ruled out for diamond. However, such fluctuations may well be present for crystals composed of atoms with large atomic numbers $Z$.

\section{Discussion and conclusions}\label{conclusions}
The instantaneous transition rate concept has been applied for channeling of particles  in straight and bent crystals. It turns out that the de-channeling length of electrons in single crystals cannot be described by a single number since the transition rate approaches a constant value only after a certain relaxation length. It is in the order of about 15 $\mu$m for our example of 855 MeV electrons channeling in (110) planes of diamond. A reasonable approach to characterize a de-channeling length would be the assignment of two numbers, the equilibrium de-channeling length and the fraction of particles for which it holds. As demonstrated above, these fractions drastically differ for the primary and secondary channeling processes.

The main deficiency of the Fokker-Planck equation may be found in the elimination of the transverse $x$ coordinate by averaging out any details associated with the dynamics and the distribution of the scattering centers across the transverse channeling coordinate. This fact reveals itself in the instantaneous drift and diffusion coefficients for each step ($\Delta x_k,\Delta z_k $), see  Eqs. (\ref{differentialDriftCoeff}) and (\ref{differentialDiffusionCoeff}). The mentioned deficiency can probably not be cured by a parameter adaption. The Fokker-Planck equation can, therefore, only be used to get a rough idea of the channeling process.

\section*{Acknowledgements} \label{Acknowledgements}
Stimulating discussions with A. V. Korol, W. Lauth, and A. V. Solov'yov are gratefully acknowledged.

This work has been financially supported by the European Union’s Horizon 2020 research
and innovation programme – the N-LIGHT project (GA 872196) within the H2020-MSCA-RISE-2019 call.

\appendix
\section*{Appendices}
\section{Double differential cross-section for scattering at electrons} \label{appendix A}
\subsection{General} \label{appendix A General}

\begin{figure}[b ]
\centering
    \includegraphics[angle=0,scale=0.65,clip]{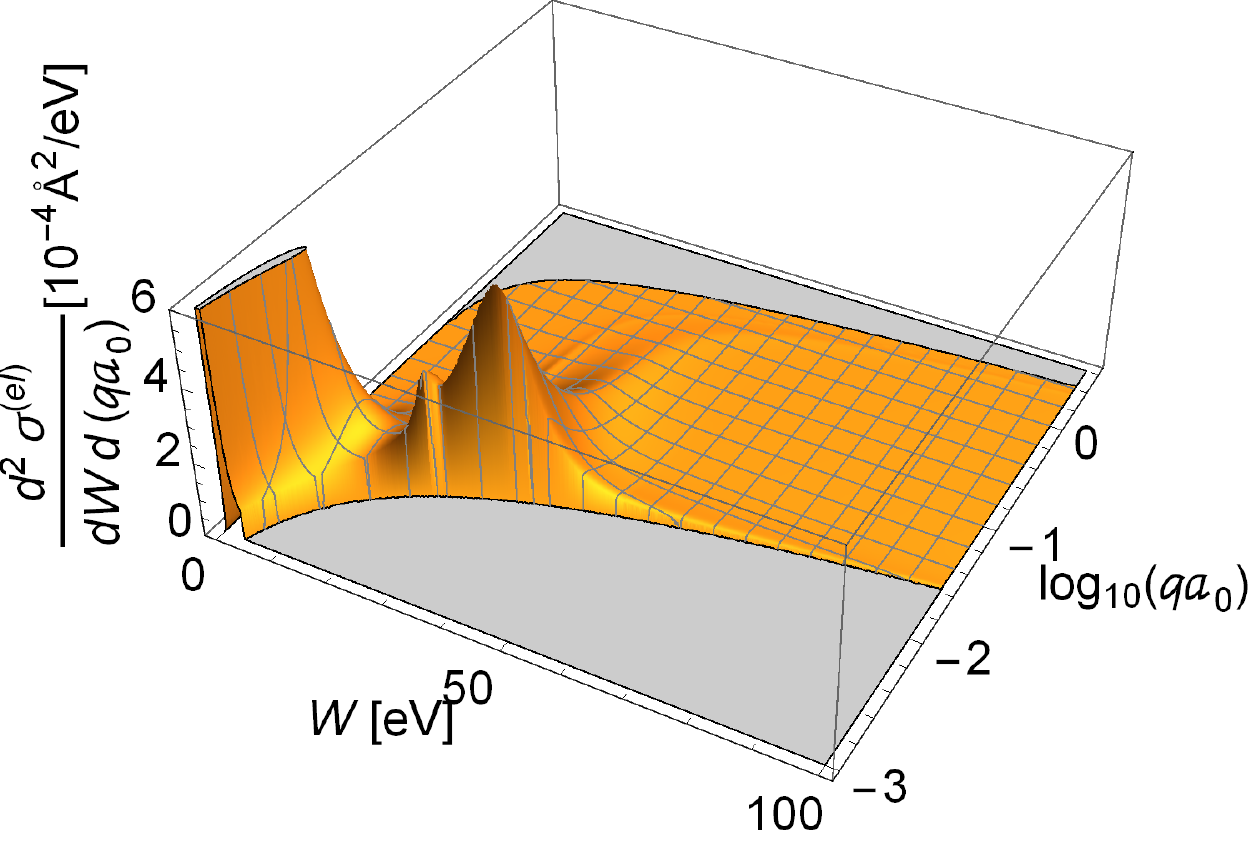}
\caption[]{Double differential cross-section according to Eq. (\ref{twodimcrossSection}) as function of the energy loss $W$ and dimensionless momentum transfer $q a_0$. The kinematical allowed region lifts out from the gray area. Two dominating features are the plasmon resonance at $W$ = 34.4 eV and $q a_0$ = 0.0096, and resonance of the transverse excitation at $W$ = 5.2 eV and $q a_0 $ = 0.005.} \label{DoubleDifferentialCrossSection}
\end{figure}

To take scattering at bound atomic electrons into account, the double differential cross-section as function of the energy transfer $W$ and the momentum transfer $q$ to a target electron is required, which can be obtained from the complex dielectric function Im$[-1/\epsilon(q,W)]$. This quantity is usually accessible only for $q$ = 0 which will be called in the following Im$[-1/\epsilon(0,W)]$ $\equiv$ Im$[-1/\epsilon(W)]$ with $\epsilon(W)=\epsilon_1(W)+\imath~\epsilon_2(W)$. To estimate Im$[-1/\epsilon(q,W)]$ from Im$[-1/\epsilon(0,W)]$ at $q$ = 0, which can be measured either by optical or electron energy-loss spectroscopy (EELS), a theoretical model is required. For longitudinal excitations the model described by Ashley \cite[and references cited therin]{Ash91} will be used. This non-relativistic model has been extended for relativistic energies by interpreting energy loss and momentum transfer in terms of relativistic quantities utilizing the papers of Inokuti \cite{Ino71} and  Fern\'{a}ndez-Varea et al. \cite{FerS05}. The contribution of the transversal cross-section play a significant role only at low momentum transfers $q a_0 < 0.02$ and low energy losses $W<$ 5 keV. It was calculated with the model of Fern\'{a}ndez-Varea et al. \cite{FerS05} and added to the longitudinal cross-section. Eq. (9) of \cite{FerS05} was used with the replacement $\epsilon(Q,W)=\epsilon_1(W)+\imath~\epsilon_2(W)$, for explanation see paragraph below Eq. (43) of \cite{FerS05}. After proper shapings the double differential cross-section reads
\begin{eqnarray}\label{twodimcrossSection}
\frac{d^2 \sigma^{(el)}}{dW dq a_0}(q a_0,W)&=&\frac{2}{\pi \beta^2 m_e c^2}\frac{1}{n_e a_0}\frac{1}{q a_0}\times
\nonumber\\
& &\hspace{-1,6 cm}
\Bigg(\frac{W-W_{min}(q a_0)}{W}\mbox{Im}\Big[\frac{-1}{\epsilon (W-W_{min}(q a_0))}\Big]+
\nonumber\\
& &\hspace{-2,2 cm}
\frac{W^2(W^2_{max}(q a_0)-W^2)\epsilon_2(W)}{[(W_{max}(q a_0)/\beta)^2-W^2~\epsilon_1(W)]^2+W^4 \epsilon^2_2(W)}\Bigg)
\end{eqnarray}
with
\begin{eqnarray}\label{inequlitiesTwodimCrossSectionW}
W_{min}(q a_0)= m_e c^2 \Big(\sqrt{1+(\alpha q a_0)^2}-1\Big) & \leq & W
\nonumber\\
\hspace{2 cm}
W\leq \beta \alpha m_e c^2 q a_0 = W_{max}(q a_0).
\end{eqnarray}
The inequalities (\ref{inequlitiesTwodimCrossSectionW}) define the relativistic correct region in which the double differential cross-section is defined for a given momentum transfer $q a_0$. All calculations above 30 keV have been done with the M{\o}ller cross sections. This is justified since well above the K shell binding energy, i.e. 0.284 keV for C, electrons can be considered as free. The function Im$[-1/\epsilon(W)]$ has been constructed with the parameters from Garcia-Molina et al. \cite[Table I]{GarA06}, and the Henke tables \cite{HenG93} for $W\leq 30$ keV else. Both distributions match at 51.2 eV. The low energy part of the double differential cross-section Eq. (\ref{twodimcrossSection}) is shown in Fig. \ref{DoubleDifferentialCrossSection}.

\begin{figure}[t b ]
\centering
    \includegraphics[angle=0,scale=0.55,clip]{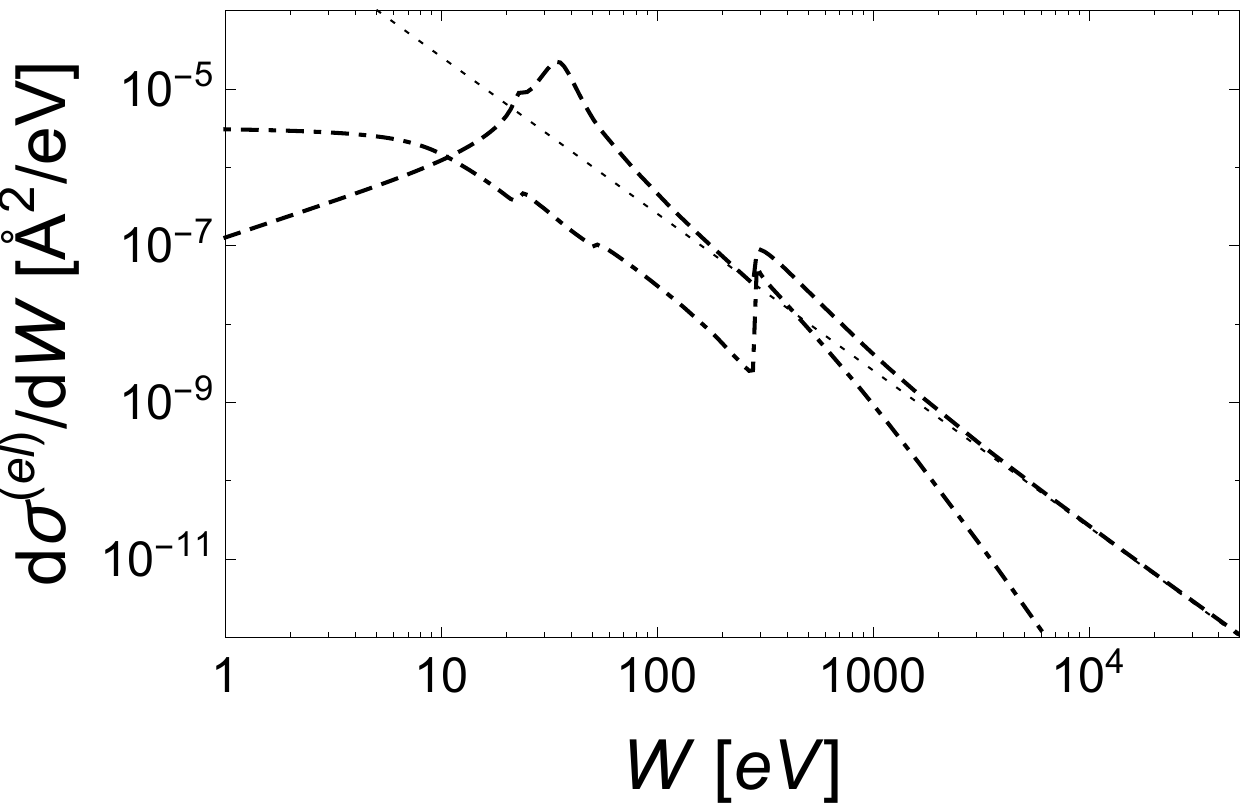}
\caption[]{Differential cross-section as function of the energy loss $W$ obtained from Eq. (\ref{energydifferentialcrossSection}). Shown are separately the contributions of the longitudinal excitation (dashed), the transverse one (dotted-dashed) and the M{\o}ller cross-section (dotted).} \label{DifferentialCrossSectionW}
\end{figure}

\begin{figure}[tbh]
\centering
    \includegraphics[angle=0,scale=0.55,clip]{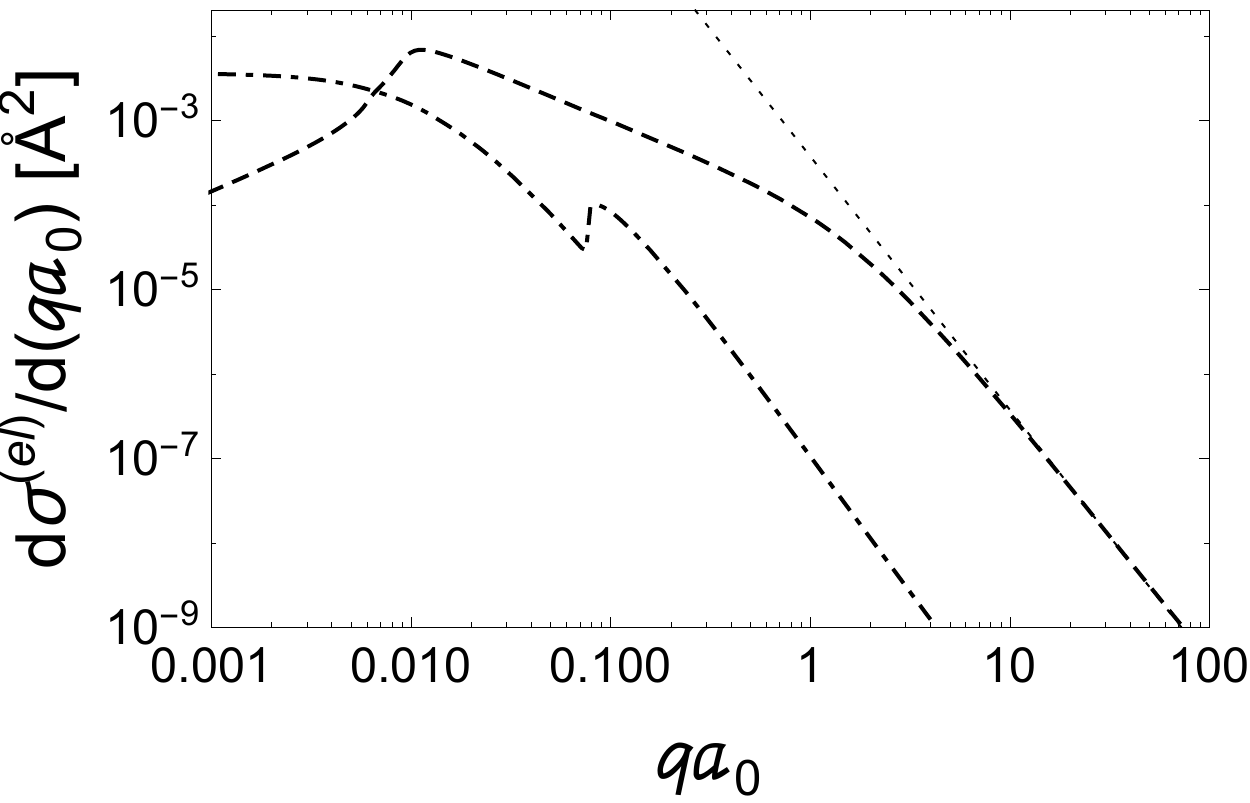}
\caption[]{Differential cross-section as function of the momentum transfer $q a_0$ obtained after integration of the double differential cross-section, Eq. (\ref{twodimcrossSection}), over the kinematical allowed energy transfer. Shown are separately the contributions of the longitudinal excitation (dashed), the transverse one (dotted-dashed) and the M{\o}ller cross-section (dotted).} \label{DifferentialCrossSectionqa0}
\end{figure}

The energy differential cross-section is obtained after integration over the kinematical allowed momentum transfer $q a_0$ at a given energy transfer $W$
\begin{eqnarray}\label{inequlitiesTwodimCrossSectionqa0}
q_{min} (W) a_0  =  W/\beta\alpha m_e c^2  \leq   q a_0
\nonumber\\
\hspace{-1 cm}
q a_0\leq \sqrt{W(W+2 m_e c^2)}\Big/\alpha m_e c^2 = q_{max}(W) a_0
\end{eqnarray}
and reads
\begin{eqnarray}\label{energydifferentialcrossSection}
\frac{d\sigma^{(el)}}{dW}(W)&=&\int\limits_{q=q_{min}(W)}^{q_{max}(W)}dq a_0~\frac{d^2 \sigma^{(el)}}{dW dq a_0}(q a_0,W)
\end{eqnarray}
This energy differential cross-section has been applied for an energy loss less than 30 keV while for higher energies the differential M{\o}ller cross-section \cite[Eq. 81.14]{BerL82} was used. Results are shown in Fig. \ref{DifferentialCrossSectionW}. At 30 keV the cross-section of Eq. (\ref{energydifferentialcrossSection}) overestimates the  M{\o}ller cross-section by 4.2\%. For the mean energy loss and the mean excitation energy one obtains $\overline{\Delta E^{(el)}/\Delta z}$ = 728.0 eV/$\mu$m and $I$ = 88.5 eV, respectively. The former deviates from the above quoted value 737.82 eV/$\mu$m, Eq. (\ref{electronMeanTransvEnergy}), by only 1.4 \%, the latter is close to the currently accepted value of 89.4 eV \cite[Table II]{TanP08}.
Similarly, the differential cross-section as function of the momentum transfer $q a_0$ is obtained after integration of Eq. (\ref{twodimcrossSection}) over the kinematical allowed energy transfer as given by Eq. (\ref{inequlitiesTwodimCrossSectionW}). Results are shown in Fig. \ref{DifferentialCrossSectionqa0}.

However, the cross-section is needed as function of the scattering angle $\theta$ rather than the momentum transfer $q a_0$. Unfortunately, it is not possible to relate the scattering angle $\theta$ uniquely to the momentum transfer $q$. It follows from relativistic kinematics that it is also a function of the energy loss $W$, see \cite[Eq. (111)]{FerS05}. Taking into account only the first term in the expansion of \cite[Eq. (105)]{FerS05}, the following relation has been deduced for $W\ll E$:
\begin{equation}
q(\theta,W)a_0 = \sqrt{\Big(\frac{\beta\gamma}{\alpha}2\sin(\theta/2)\Big)^2+\Big( \frac{W}{\beta\alpha~ m_e c^2}\Big)^2}.
\label{replacementqa0}
\end{equation}
After some algebraic manipulations one obtains for the double differential cross-section in terms of the scattering angle $\theta$ and energy loss $W$
\begin{eqnarray}\label{twodimcrossectionomega}
\frac{d^2 \sigma^{(el)}}{dW d\Omega}(\theta,W)&=&\Big(\frac{\beta\gamma}{\alpha}\Big)^2 \frac{1}{2\pi q(\theta,W)a_0}\times
\nonumber\\
& &\hspace{0,4 cm}
\frac{d^2 \sigma^{(el)}}{dW d q a_0}(q a_0,W)\Big|_{q a_0 \rightarrow q(\theta,W)a_0 }
\end{eqnarray}
within the kinematical allowed region
\begin{eqnarray}
W_{min}(\theta)=\frac{(\beta\gamma)^2 \sin^2\theta}{2+(\gamma-1)\sin^2\theta}m_e c^2\leq W\leq\frac{\gamma - 1}{2} m_e c^2 \label{inequlitiesTwodimCrossSectionThetaWW} ~~~~~~~\\
0\leq\sin^2 \theta \leq \frac{2 W}{(\gamma-1)((\gamma+1)m_e c^2-W)}=\sin^2 \theta_{max}(W).~~ \label{inequlitiesTwodimCrossSectionThetaWTheta}
\end{eqnarray}
The lefthand side of Eq. (\ref{inequlitiesTwodimCrossSectionThetaWW}) and the righthand side of Eq. (\ref{inequlitiesTwodimCrossSectionThetaWTheta}) follow from kinematics for the M{\o}ller scattering.

\subsection{The role of resonances} \label{appendix A Discussion}
A few remarks will be added on the role of the strong plasmon resonance seen in Fig. \ref{DoubleDifferentialCrossSection}. The question may arise what are possible implications if electrons are treated as a free gas and M{\o}ller scattering is applied. Since the M{\o}ller cross-section diverges at zero energy transfer, a proper low energy cut-off must be chosen. In case the total electron-electron scattering cross-section is unknown, one is left with the formidable task how to get it. Supposing the cross-section 5.576 $\cdot 10^{-4}~\AA^2$ would be known, than the cut-off energy can be determined to be 4.579 eV.
\begin{figure}[tbh]
\centering
    \includegraphics[angle=0,scale=0.55,clip]{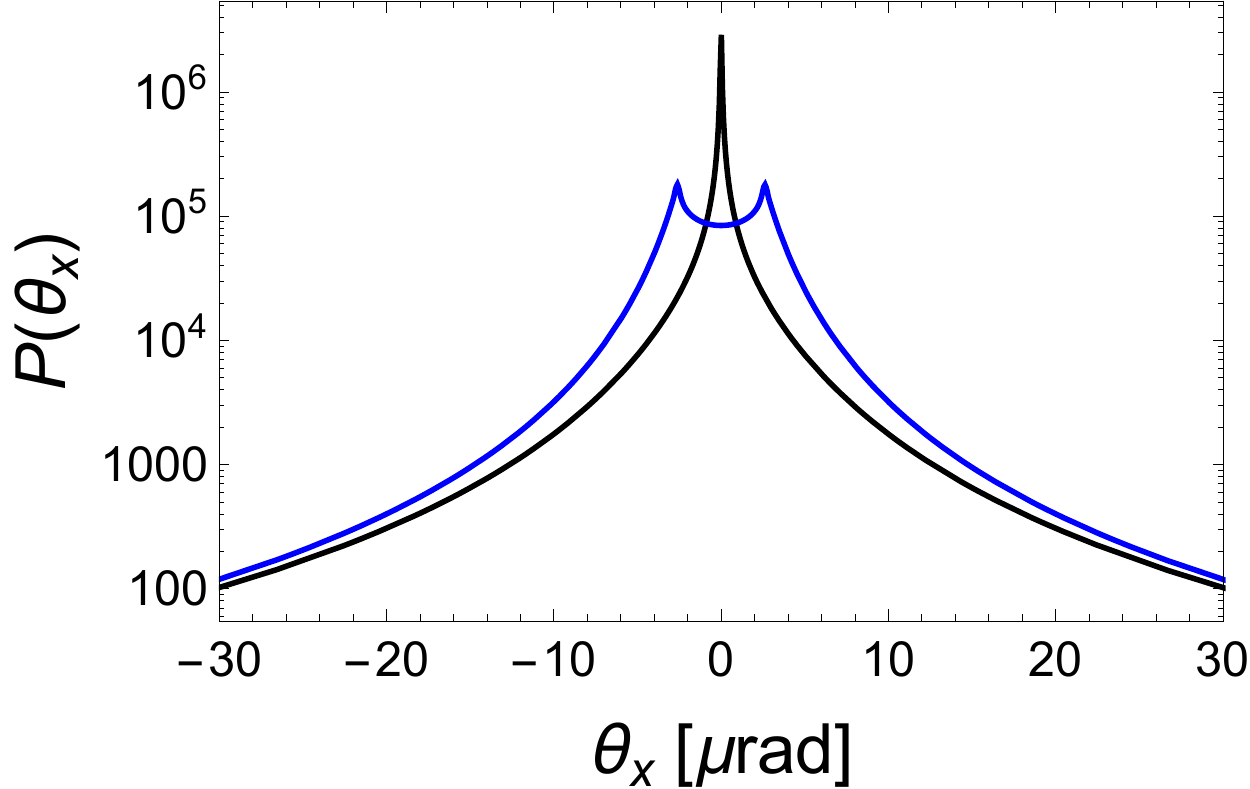}
\caption[]{Normalized electron-electron scattering distributions for carbon. For the black curve resonances are taken into account, for the blue only M{\o}ller scattering with a low energy cut-off energy of 4.579 eV yielding the same cross-section.} \label{ScatteringProbabilitiesEl}
\end{figure}
In Fig. \ref{ScatteringProbabilitiesEl} the impact on the scattering distribution is shown. Neglecting the resonances leads to a broadening of the scattering distribution for small angles.  Although the effect looks quite significant, the mean transverse energy increases by only 17\%. The reason is that the probability density deviates essentially at low scattering angles. For an inferred cut-off at the free atom 2p binding energy of 11.26 eV, the scattering distribution broadens also for rather large angles. Nevertheless, the mean transverse energy increases also only by a small amount. The reason is that the effect of the broadening is compensated by the reduced cross-section, which is a factor a factor of 2.46 smaller, resulting in less collisions per unit path length. However, the fluctuations of the transverse energy transfer will increase.

While it appears that the rather complicated treatment of the inelastic electron-electron interaction is not worth the effort in view of the small effect, this is not the case for channeling of positrons for which the positron-electron interaction dominates de-channeling. The formalism is the same as described here. Only the M{\o}ller cross-section must be replaced by the Bhabha one. This will be the subject of a forthcoming paper.

\section{Derivation of the transverse energy transfer at a collision} \label{appendix B}

\begin{figure}[tbh]
\centering
    \includegraphics[angle=0,scale=0.32,clip]{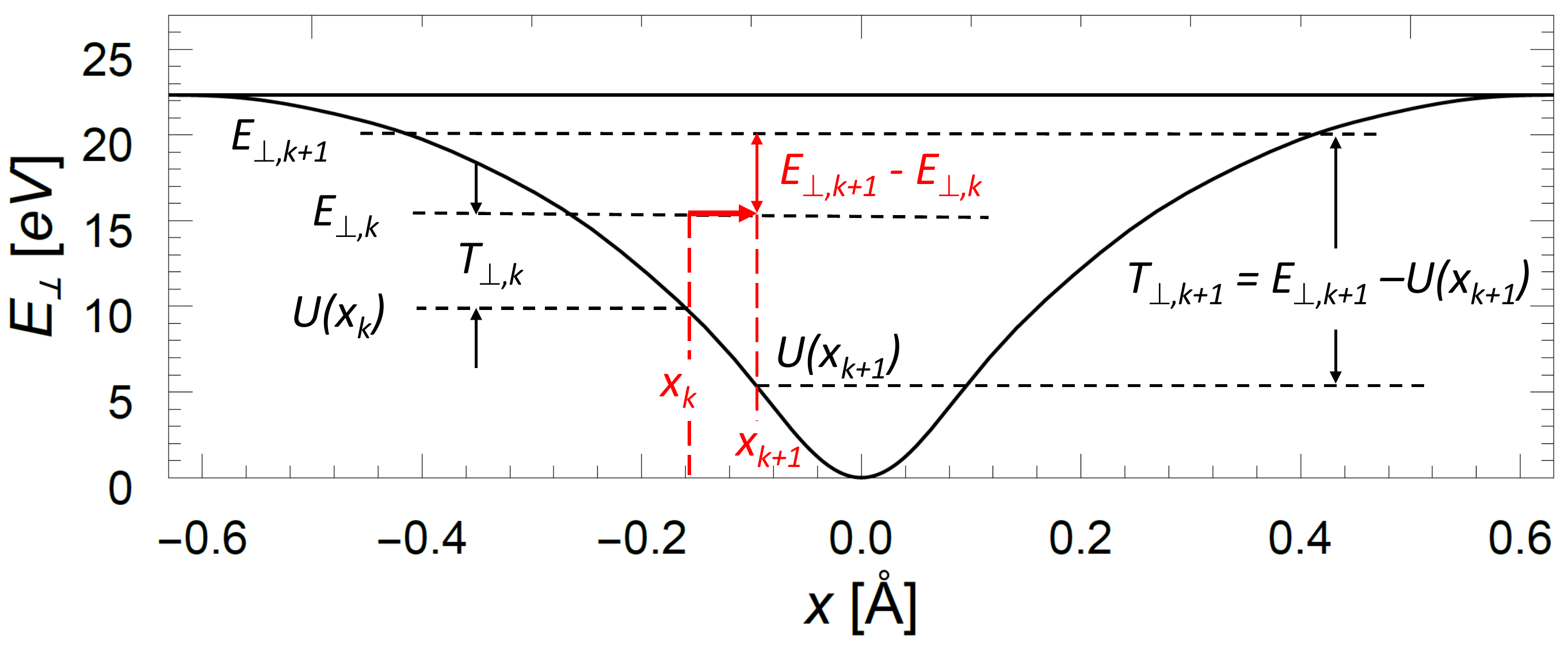}
\caption[]{For explanation how Eq. (\ref{NewEperp}) has been derived in the potential of the (110) plane of diamond. In a step from $x_{k}\rightarrow x_{k+1}$, indicated by red colour, the scattering angle may change by $\Delta \vartheta_k$ which results in the new energy $E_{\bot,k+1}$ according to Eq. (\ref{NewEperp}). In the numerical simulation the digitization interval is $\Delta x \cong$ 0.01 {\AA}.} \label{AppendixDeltaEperp}
\end{figure}

In the following it will be explained how the Eqs. (\ref{newScatteringAngle}) and (\ref{NewEperp}) come about. The first term of Eq. (\ref{newScatteringAngle}) is simply the angle as derived from the general relation for the kinetic energy $T_{\perp,{k}} = p v/2\cdot\vartheta_{k}^2$. For the second equation we refer to Fig. \ref{AppendixDeltaEperp}. At the step from $x_k\rightarrow x_{k+1}$ the kinetic energy changes due to the potential energy difference $U(x_{k})-U(x_{k+1})$ as well as due to a possible scattering by the angle $\Delta  \vartheta_k$. A change of the transverse kinetic energy results from $T_{\bot,k} = E_{\bot,k}-U(x_{k}) = pv/2\cdot \vartheta_{k}^2$ to $T_{\bot,k+1} = E_{\bot,k+1}-U(x_{k+1}) =U(x_{k})-U(x_{k+1})+ pv/2\cdot(\vartheta_{k}+\Delta  \vartheta_k)^2$. Evaluating the latter two equations for $E_{\bot,k+1}$ results in Eq. (\ref{NewEperp}). Special care is required if the particle is reflected at the potential borders.

\section{Examples of channeling trajectories} \label{appendix C}
\begin{figure}[tbh]
\centering
    \includegraphics[angle=0,scale=0.60,clip]{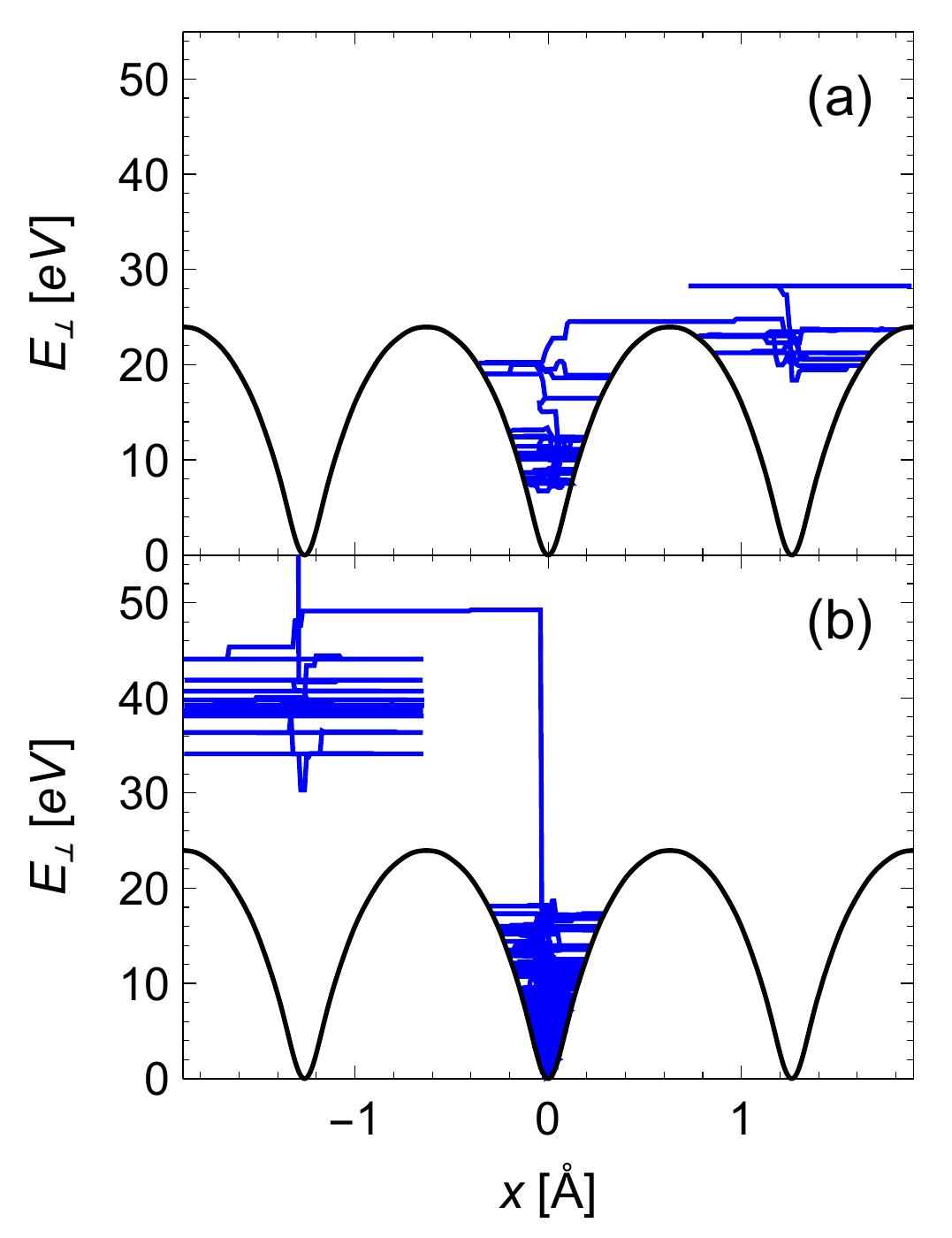}
\caption[]{Examples of channeling trajectories. Panel (a) shows an example with re-channeling and subsequent de-channeling. The electron leaves the right potential pocket to the right. Panel (b) shows an example of a long channeling history in the primary potential pocket. The de-channeled electron never re-channels and leaves the canvas at the maximum energy, indicated by the second vertical jump.} \label{ChannelingTrajectories}
\end{figure}

In Fig. \ref{ChannelingTrajectories} typical examples of trajectories are shown in the $(E_\perp, x)$ plane. The particle is originally captured in the middle potential pocket. Only three potential pockets are shown.  Large changes of the transverse energy happen essentially only at an overlap with the atomic density at the potential minimum, smaller ones also outside by collisions with electrons. 


\bibliographystyle{spphys}
\bibliography{bibfileBa}

\end{document}